\documentclass{llncs}

\usepackage{xcolor}
\usepackage{amsmath,amssymb,amsfonts}
\usepackage{mathtools}
\usepackage{xspace}
\usepackage{hyperref}
\usepackage{subcaption}
\usepackage{paralist}
\usepackage{microtype}
\usepackage{wrapfig}
\usepackage{colonequals}
\usepackage{tikz}
\usepackage{cite}
\usepackage{multirow}
\usetikzlibrary{arrows.meta,decorations.pathmorphing,positioning,fit,trees,shapes,shadows,automata,calc,decorations.markings,patterns,arrows} 
\tikzset{>=latex}
\usepackage{pgfplots}
\pgfplotsset{compat=1.15}
\tikzset{
  dot hidden/.style={},
  line hidden/.style={},
  dice hidden/.style={},
  dot color/.style={dot hidden/.append style={color=#1}},
  dot color/.default=black,
  line color/.style={line hidden/.append style={color=#1}},
  line color/.default=black,
  dice color/.style={dice hidden/.append style={color=#1,fill}},
  dice color/.default=white
}\def\dotsize{0.1}
\newcommand{\drawdie}[2][]{%
\begin{tikzpicture}[x=0.9em,y=0.9em,#1]
  \draw 	[thick, rounded corners=0.5,line hidden,dice hidden] (0,0) rectangle (1,1);
  \ifodd#2
    \fill[dot hidden] (0.5,0.5) circle (\dotsize);
  \fi
  \ifnum#2>1
  \fill[dot hidden] (0.25,0.25) circle (\dotsize);
  \fill[dot hidden] (0.75,0.75) circle (\dotsize);
  \ifnum#2>3
    \fill[dot hidden] (0.25,0.75) circle (\dotsize);
    \fill[dot hidden] (0.75,0.25) circle (\dotsize);
    \ifnum#2>5
      \fill[dot hidden] (0.75,0.5) circle (\dotsize);
      \fill[dot hidden] (0.25,0.5) circle (\dotsize);
    \fi
  \fi
\fi
\end{tikzpicture}%
}

\usepackage{algorithmicx,algorithm}
\usepackage[noend]{algpseudocode}
\usepackage{nicefrac}

\renewcommand{\emptyset}{\varnothing}
\renewcommand{\leq}{\leqslant}
\renewcommand{\geq}{\geqslant}

\usepackage{listings}

\include{prismstyle}

\newcommand{\Distr}{\mathit{Distr}}

\newcommand{\R}{\mathbb{R}}
\newcommand{\Q}{\mathbb{Q}}

\newcommand{\N}{\mathbb{N}}

\newcommand{\sinit}{s_{\mathit{I}}} 

\newcommand{\dtmc}{D}

\newcommand{\mdp}{M}

\newcommand{\act}[1][a]{\alpha} 
\newcommand{\Act}{\mathit{Act}}        

\newcommand{\probdtmc}{\probmdp}
\newcommand{\pdtmc}{\mathcal{\dtmc}}

\newcommand{\probmdp}{\mathcal{P}}

\newcommand{\rel}[2][]{\ensuremath{\textsf{rel}_{#1}(#2)}}

\newcommand{\ie}{i.\,e.}

\newcommand{\tool}[1]{\texttt{#1}\xspace}
\newcommand{\prophesy}{\tool{PROPhESY}}

\DeclareRobustCommand{\cpp}
{\valign{\vfil\hbox{##}\vfil\cr
   \textsf{C\kern-.1em}\cr
   $\hbox{\fontsize{\sf@size}{0}\textbf{+\kern-0.05em+}}$\cr}%
}

\selectcolormodel{cmy}

\newcommand{\commentside}[2]{\marginpar{\color{#1}#2}}

\newif\ifdebug
\debugtrue

\ifdebug
 \newcommand{\nils}[1]{\color{red}#1\color{black}\xspace}
 
 \newcommand{\nj}[1]{\commentside{teal}{\tiny{N: #1} }}
 \newcommand{\sj}[1]{\commentside{red}{\tiny{SJ: #1} }}
 \newcommand{\cd}[1]{\commentside{blue}{\tiny{CD: #1} }}
 
 \newcommand{\mv}[1]{\commentside{green}{\tiny{MV: #1} }}

\else
 \newcommand{\nils}[1]{}
\newcommand{\nj}[1]{}
\newcommand{\sj}[1]{}
\newcommand{\fc}[1]{}
\newcommand{\cd}[1]{}
\newcommand{\mv}[1]{}
\fi

\tikzset{cross/.style={cross out, draw=black, minimum size=2*(#1-\pgflinewidth), inner sep=0pt, outer sep=0pt},
cross/.default={1pt}}

\usetikzlibrary{patterns}

\newcommand{\distsonesthree}{1.5cm}
\newcommand{\distforsone}{1.5cm}

\begin{document}

\title{Parameter Synthesis in Markov Models: \\ A Gentle Survey\thanks{This work has been supported by the DFG RTG 2236 ``UnRAVeL'' and the ERC Advanced Grant 787914 ``FRAPPANT''.}}
\author{Nils Jansen\inst{1}, Sebastian Junges\inst{1} and Joost-Pieter Katoen\inst{2}}
\institute{
Radboud University, Nijmegen, The Netherlands
\and
RWTH Aachen University, Aachen, Germany 
}
\titlerunning{Parameter Synthesis in Markov Models: A Survey}
\authorrunning{Junges and Katoen}

\maketitle
\thispagestyle{plain}\pagestyle{plain}  

\begin{abstract}
This paper surveys the analysis of parametric Markov models whose transitions are labelled with functions over a finite set of parameters.
These models are symbolic representations of uncountable many concrete probabilistic models, each obtained by instantiating the parameters.
We consider various analysis problems for a given logical specification $\varphi$:
do all parameter instantiations within a given region of parameter values satisfy 
$\varphi$?, which
instantiations satisfy $\varphi$ and which ones do not?, and how can all such 
instantiations be characterised, either exactly or approximately? 
We address theoretical complexity results and describe the main ideas underlying state-of-the-art algorithms that established an impressive leap over the last decade enabling the fully automated analysis of models with millions of states and thousands of parameters.
\end{abstract}

\section{Introduction}
Markov models are ubiquitous.
Markov chains (MCs) are central in performance and dependability analysis, whereas Markov decision processes (MDPs) are key in stochastic decision making and planning in AI.
A standard assumption in these models is that all probabilities are precisely known.
This assumption is often too severe.
System quantities such as component fault rates, molecule reaction rates, packet loss ratios, etc.\ are often not, or at best partially, known. 
\begin{wrapfigure}[11]{r}{0.36\textwidth}
\vspace*{-0.1cm}
\centering
\includegraphics[scale=0.28]{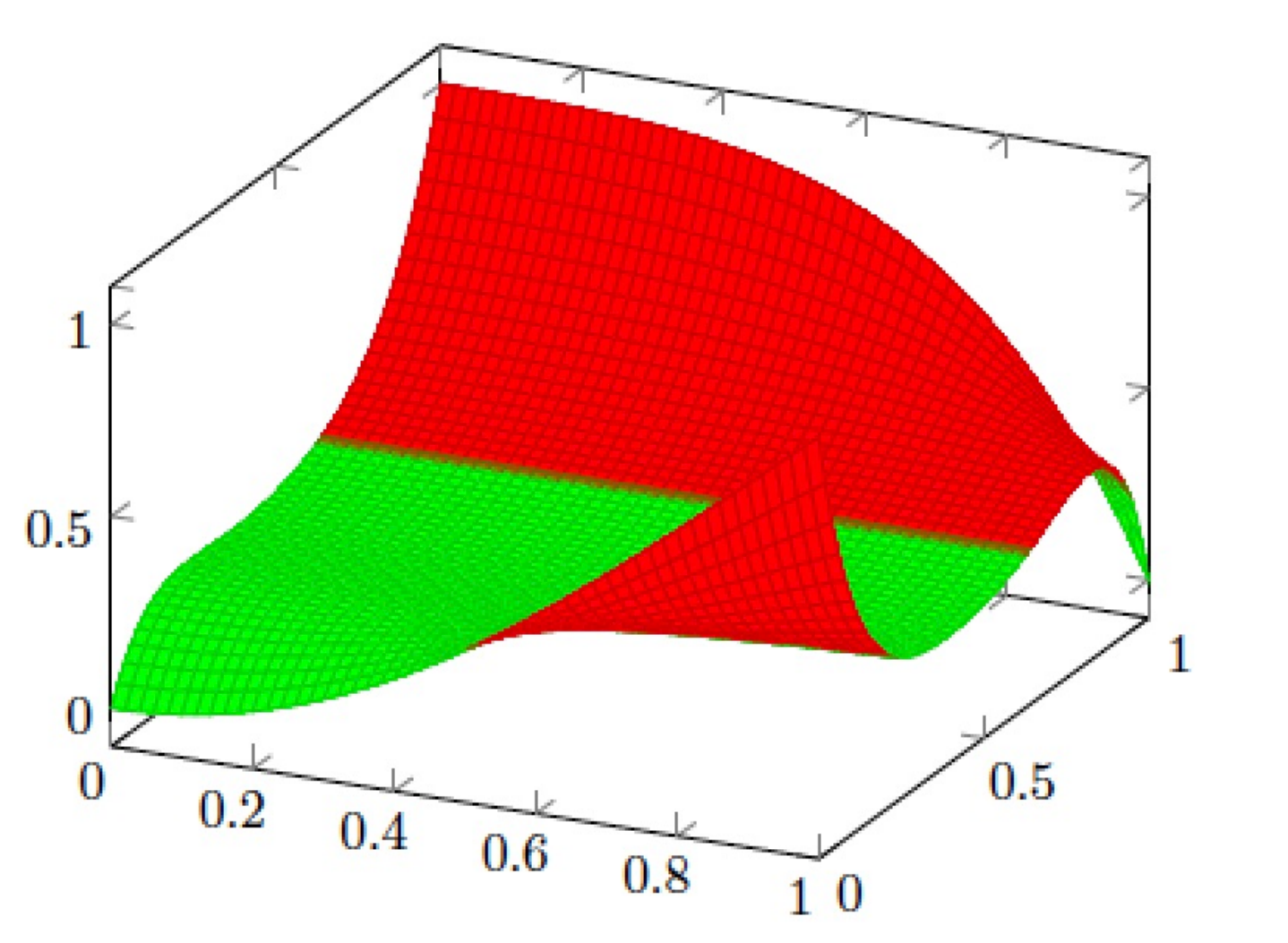}
\vspace{-0.3cm}
\caption{For which gate failure rates is the circuit reliable?}
\label{fig:plot-nand}
\end{wrapfigure}
\paragraph{A motivating example.}
In early stages of reliable system design, the concrete failure rate of components~\cite{Cousineau2009} is deliberately left unspecified.
Let us illustrate this by considering a multiplexed NAND integrated circuit.
As such circuits are built at ever smaller scale, they are prone to defects and/or to exhibit transient failures.
The analysis of a parametric Markov chain model of the NAND circuit~\cite{DBLP:journals/pieee/HaselmanH10} can provide insights into the effect of failure probabilities of the gates on the overall circuit's reliability.
The unknown random gate failure rates are the model's parameters. 
Questions of interest are e.g., for which gate failure rates are at least 70\% of the NAND outputs correct, or which failure rates minimise the circuit's reliability?
More advanced objectives are to determine either exactly or approximately \emph{all} possible failure rates such that the NAND's failure probability is below $\nicefrac{3}{10}$, or even to provide a closed-form formula of how the circuit's reliability depends on the vulnerability of its gates.
Fig.~\ref{fig:plot-nand}, for instance, plots exact synthesis results (obtained in a few minutes) for two gates with unknown failure rate ($x$- and $y$-axis) indicating all gate's failure rates leading to circuit failure probabilities ($z$-axis) above (red) and below (green) the threshold $\nicefrac{3}{10}$.

\paragraph{Aims of this paper.}
Our aim is to provide insight in the underlying algorithmic techniques and the complexity to answer the above questions of interest.
We do so by surveying the automated analysis of parametric Markov models.
For the sake of simplicity, we consider \emph{parametric Markov chains (pMCs)} and do not consider parametric versions of MDPs.
Yet, various of the presented techniques are also applicable to non-deterministic models.
Parametric MCs are classical discrete-time Markov chains where the transition probabilities are specified by polynomials over real-valued parameters, e.g., $x$, $1{-}x$, and $1{-}x{\cdot y}{+}z$.
Intervals over parameters are simple instances imposing constant lower and upper bounds on each parameter~\cite{DBLP:journals/rc/KozineU02,DBLP:journals/ai/GivanLD00}.
The setting here is more liberal as it includes the possibility to express parameter dependencies: a parameter can occur at several transitions. 
This often leads to solving trade-offs: for some states it may beneficial to have a small value of parameter $x$, whereas for other states a large value of $x$ is better, e.g., to increase the probability to reach a certain target state.

\paragraph{Parameter synthesis problems.}
We consider various problems for parametric MCs and a given logical specification $\varphi$: (1) do all parameter instantiations, within a given region of parameter values, satisfy $\varphi$?, (2) which instantiations satisfy $\varphi$ and which ones do not?, and (3) how can all such instantiations be characterised, either exactly or approximately? 
These questions are intrinsically hard: parameters can take uncountable many different values and the parameters may depend on each other.
We address theoretical complexity results and describe the main ideas underlying state-of-the-art algorithms for (1) through (3).
For the sake of simplicity, we focus on specifications $\varphi$ that impose thresholds on infinite-horizon reachability probabilities; e.g., can a bad state be reached with a probability at most 10$^{-6}$? 
The presented algorithms can however also be generalised in a straightforward manner to total expected reward objectives such as: is the expected cost to reach a goal state at most 10$^3$? 
This survey aims to give an insight into the main advances during the last decade. 
Whereas initial algorithms could only handle a handful of parameters\footnote{And this was viewed by quite some scepticism, e.g., our initial paper on this topic received reviews saying ``this is a fun problem that is unlikely to ever result in practical advancements.''}, these advancements enable \emph{the fully automated analysis of models with millions of states and thousands of parameters}.

\paragraph{POMDP controller synthesis.}
Parametric MCs can be seen as Markov chains with partial information about the transition probabilities.
Indeed, these transition probabilities are solutions of functions over the parameters.
As the choice of the parameter values is assumed to be fully non-deterministic (these values are e.g., not subject to a probability distribution), pMCs can be seen as a kind of MDP with partial information.
This raises the question whether there is a relation to the well-known model of partially-observable MDPs (POMDPs)~\cite{Kochenderfer2015,DBLP:books/daglib/0023820,DBLP:books/sp/12/Spaan12}.
POMDPs are a prominent model in sequential decision-making under uncertainty: decisions have to be made under partial knowledge about the environment.
Intuitively, POMDP policies (aka: controllers) have to make optimal decisions based on partial information about the visited states.
Indeed, there exists a direct link between parameter synthesis for pMCs and controller synthesis for POMDPs.
In fact, finding a finite-memory policy that guarantees the satisfaction of a reachability property $\varphi$ in a POMDP is equivalent to feasibility --- find a parameter instantiation satisfying $\varphi$ --- in a corresponding pMC.

\paragraph{Highlights.}
Section~\ref{sec:complexity} shows that the decision problem of feasibility, does there exist a parameter instantiation such that specification $\varphi$ is satisfied, is ETR-complete.
In particular, the fact that ETR-satisfiability problems can be encoded as reachability problems in pMCs is of interest.
Section~\ref{sec:feasibility} describes how convex optimisation can be used to solve feasibility approximatively and how a tight integration with model checking improves this further.
Section~\ref{sec:parameter-lifting} describes an abstraction-based approach that reduces checking whether a region of parameter instantiations satisfies $\varphi$ on a pMC to a parameter-less model checking problem on an MDP.
Section~\ref{sec:pomdps} details the connection between feasibility in pMCs to POMDP controller synthesis.
This link closes a complexity gap in POMDP controller synthesis, and enables using parameter synthesis algorithms such as the integrated model checking--convex optimisation for POMDP controller synthesis.
Almost all sections include indications about the kind of synthesis problems (in terms of model size, number of parameters, and precision) that can be solved using current algorithms and software tools.

\section{Parametric Markov Chains}
\label{sec:preliminaries}

Let $X$ be a set of $n$ real-valued parameters (or variables) $x_1, \ldots, x_n$.
The parameters $x_i$ should be considered as symbols.
Their values are defined by a parameter instantiation, i.e., a function $v \colon X \to \R$.
The set $V \subseteq \R^n$ of all parameter values is called the parameter space.
A set $R \subseteq V$ of instantiations is a \emph{region}.

Let $\Q[X]$ denote the set of multivariate polynomials over $X$ with rational coefficients.
A polynomial $f \in \Q[X]$ can be interpreted as a function $f \colon \R^n \to \R$ where $f(v)$ is obtained by replacing each occurrence of $x_i$ in $f$ by $v(x_i)$; e.g., for $f = 2x_1{\cdot}x_2+x_1^2$ with $v(x_1) = 2$ and $v(x_2) = 3$, we have $f(v) = 16$. 
To make clear where substitution occurs, we write $f[v]$ instead of $f(v)$ from now on.

\begin{definition}\label{def:pmc}
A pMC $\pdtmc$ is a tuple $(S,\sinit,X,\probdtmc)$ with a finite set $S$ of \emph{states}, an \emph{initial state} $\sinit \in S$, a finite set $X$ of real-valued variables \emph{(parameters)} and a \emph{transition function} $\probdtmc \colon S\times S \to \Q[X]$.
\end{definition}

\begin{figure}[t]
\begin{subfigure}{0.32\textwidth}
\centering
\scalebox{0.75}{
\begin{tikzpicture}[scale=1, die/.style={inner sep=0,outer sep=0}, every node/.style={font=\scriptsize}, st/.style={draw, circle, inner sep=2pt, minimum size=15pt}]
    
    \node [st,fill=gray!50] (s0) at (0,0) {$s_0$};
    \node [] (leftdummy)  [on grid, left=1.2cm of s0] {};
    \node [] (rightdummy) [on grid, right=1.2cm of s0] {};
    \node [st] (s1) [on grid, below=1.1cm of leftdummy] {$s_1$};
    \node [st] (s2) [on grid, below=1.1cm of rightdummy] {$s_2$};
    \node [st,fill=gray!50] (s3) [on grid, below=1.3cm of s1, xshift=-0.5cm] {$s_3$};
        \node [st,fill=gray!50] (s4) [on grid, below=1.3cm of s1, xshift=0.5cm] {$s_4$};
    \node [st,fill=gray!50] (s5) [on grid, below=1.3cm of s2, xshift=-0.5cm] {$s_5$};
    
    \node [st,fill=gray!50] (s6) [on grid, below=1.3cm of s2, xshift=0.5cm] {$s_6$};

    \node[die, scale=2, below=0.6cm of s3, xshift=-0.1cm] (X1) {\drawdie{1}};
    \node[die, scale=2, right=0.26cm of X1, inner sep=0pt] (X2) {\drawdie{2}};
    \node[die, scale=2, right=0.26cm of X2] (X3) {\drawdie{3}};
    \node[die, scale=2, right=0.26cm of X3] (X4) {\drawdie{4}};
    \node[die, scale=2, right=0.26cm of X4] (X5) {\drawdie{5}};
    \node[die, scale=2, right=0.26cm of X5] (X6) {\drawdie{6}};
    
    \draw ($(s0)-(0.7,0)$) edge[->] (s0);
    \draw (s0) edge[->] node[right] {\scriptsize$\nicefrac{1}{2}$} (s1);
    \draw (s0) edge[->] node[right] {\scriptsize$\nicefrac{1}{2}$} (s2);
    \draw (s1) edge[bend left, ->] node[left] {\scriptsize$\nicefrac{1}{2}$} (s3);
    \draw (s1) edge[->] node[right] {\scriptsize$\nicefrac{1}{2}$} (s4);
    \draw (s3) edge[bend left, ->] node[left] {\scriptsize$\nicefrac{1}{2}$} (s1);
    \draw (s3) edge[->] node[left] {\scriptsize$\nicefrac{1}{2}$} (X1);
    \draw (s4) edge[->] node[left] {\scriptsize$\nicefrac{1}{2}$} (X2);
    \draw (s4) edge[->] node[right] {\scriptsize$\nicefrac{1}{2}$} (X3);
    \draw (s2) edge[bend left, ->] node[left] {\scriptsize$\nicefrac{1}{2}$} (s5);
    \draw (s2) edge[->] node[right] {\scriptsize$\nicefrac{1}{2}$} (s6);
    \draw (s5) edge[bend left, ->] node[left] {\scriptsize$\nicefrac{1}{2}$} (s2);
    \draw (s5) edge[->] node[left] {\scriptsize$\nicefrac{1}{2}$} (X4);
    \draw (s6) edge[->] node[left] {\scriptsize$\nicefrac{1}{2}$} (X5);
    \draw (s6) edge[->] node[right] {\scriptsize$\nicefrac{1}{2}$} (X6);
    
    \node[draw=white, rectangle, fit=(current bounding box)] {};
\end{tikzpicture}
}
\subcaption{fair coins}
\label{fig:kydie}	
\end{subfigure}
\begin{subfigure}{0.32\textwidth}
\centering
\scalebox{0.75}{
\begin{tikzpicture}[scale=1, die/.style={inner sep=0,outer sep=0}, every node/.style={font=\scriptsize}, st/.style={draw, circle, inner sep=2pt, minimum size=15pt},baseline=(s0)]
    
    \node [st,fill=gray!50] (s0) at (0,0) {$s_0$};
    \node [] (leftdummy)  [on grid, left=1.2cm of s0] {};
    \node [] (rightdummy) [on grid, right=1.2cm of s0] {};
    \node [st] (s1) [on grid, below=1.1cm of leftdummy] {$s_1$};
    \node [st] (s2) [on grid, below=1.1cm of rightdummy] {$s_2$};
    \node [st,fill=gray!50] (s3) [on grid, below=1.3cm of s1, xshift=-0.5cm] {$s_3$};
        \node [st,fill=gray!50] (s4) [on grid, below=1.3cm of s1, xshift=0.5cm] {$s_4$};
    \node [st,fill=gray!50] (s5) [on grid, below=1.3cm of s2, xshift=-0.5cm] {$s_5$};
    
    \node [st,fill=gray!50] (s6) [on grid, below=1.3cm of s2, xshift=0.5cm] {$s_6$};

    \node[die, scale=2, below=0.6cm of s3, xshift=-0.1cm] (X1) {\drawdie{1}};
    \node[die, scale=2, right=0.26cm of X1, inner sep=0pt] (X2) {\drawdie{2}};
    \node[die, scale=2, right=0.26cm of X2] (X3) {\drawdie{3}};
    \node[die, scale=2, right=0.26cm of X3] (X4) {\drawdie{4}};
    \node[die, scale=2, right=0.26cm of X4] (X5) {\drawdie{5}};
    \node[die, scale=2, right=0.26cm of X5] (X6) {\drawdie{6}};
    
    \draw ($(s0)-(0.7,0)$) edge[->] (s0);
    \draw (s0) edge[->] node[right] {\scriptsize$\nicefrac{2}{5}$} (s1);
    \draw (s0) edge[->] node[right] {\scriptsize$\nicefrac{3}{5}$} (s2);
    \draw (s1) edge[bend left, ->] node[left] {\scriptsize$\nicefrac{7}{10}$} (s3);
    \draw (s1) edge[->] node[right] {\scriptsize$\nicefrac{3}{10}$} (s4);
    \draw (s3) edge[bend left, ->] node[left] {\scriptsize$\nicefrac{2}{5}$} (s1);
    \draw (s3) edge[->] node[left] {\scriptsize$\nicefrac{3}{5}$} (X1);
    \draw (s4) edge[->] node[left] {\scriptsize$\nicefrac{3}{5}$} (X2);
    \draw (s4) edge[->] node[right] {\scriptsize$\nicefrac{2}{5}$} (X3);
    \draw (s2) edge[bend left, ->] node[left] {\scriptsize$\nicefrac{7}{10}$} (s5);
    \draw (s2) edge[->] node[right] {\scriptsize$\nicefrac{3}{10}$} (s6);
    \draw (s5) edge[bend left, ->] node[left] {\scriptsize$\nicefrac{2}{5}$} (s2);
    \draw (s5) edge[->] node[left] {\scriptsize$\nicefrac{3}{5}$} (X4);
    \draw (s6) edge[->] node[left] {\scriptsize$\nicefrac{3}{5}$} (X5);
    \draw (s6) edge[->] node[right] {\scriptsize$\nicefrac{2}{5}$} (X6);
    
    \node[draw=white, rectangle, fit=(current bounding box)] {};
\end{tikzpicture}
}
\subcaption{biased coins}
\label{fig:intro:ky:biased}	
\end{subfigure}
\begin{subfigure}{0.32\textwidth}
\centering
\scalebox{0.75}{
\begin{tikzpicture}[scale=1, die/.style={inner sep=0,outer sep=0}, every node/.style={font=\scriptsize}, st/.style={draw, circle, inner sep=2pt, minimum size=15pt},baseline=(s0)]
    
    \node [st,fill=gray!50] (s0) at (0,0) {$s_0$};
    \node [] (leftdummy)  [on grid, left=1.2cm of s0] {};
    \node [] (rightdummy) [on grid, right=1.2cm of s0] {};
    \node [st] (s1) [on grid, below=1.1cm of leftdummy] {$s_1$};
    \node [st] (s2) [on grid, below=1.1cm of rightdummy] {$s_2$};
    \node [st,fill=gray!50] (s3) [on grid, below=1.3cm of s1, xshift=-0.5cm] {$s_3$};
        \node [st,fill=gray!50] (s4) [on grid, below=1.3cm of s1, xshift=0.5cm] {$s_4$};
    \node [st,fill=gray!50] (s5) [on grid, below=1.3cm of s2, xshift=-0.5cm] {$s_5$};
    
    \node [st,fill=gray!50] (s6) [on grid, below=1.3cm of s2, xshift=0.5cm] {$s_6$};

    \node[die, scale=2, below=0.6cm of s3, xshift=-0.1cm] (X1) {\drawdie{1}};
    \node[die, scale=2, right=0.26cm of X1, inner sep=0pt] (X2) {\drawdie{2}};
    \node[die, scale=2, right=0.26cm of X2] (X3) {\drawdie{3}};
    \node[die, scale=2, right=0.26cm of X3] (X4) {\drawdie{4}};
    \node[die, scale=2, right=0.26cm of X4] (X5) {\drawdie{5}};
    \node[die, scale=2, right=0.26cm of X5] (X6) {\drawdie{6}};
    
    \draw ($(s0)-(0.7,0)$) edge[->] (s0);
    \draw (s0) edge[->] node[right] {\scriptsize$x$} (s1);
    \draw (s0) edge[->] node[right] {\scriptsize$1-x$} (s2);
    \draw (s1) edge[bend left, ->] node[left] {\scriptsize$y$} (s3);
    \draw (s1) edge[->] node[right] {\scriptsize$1-y$} (s4);
    \draw (s3) edge[bend left, ->] node[left] {\scriptsize$x$} (s1);
    \draw (s3) edge[->] node[left] {\scriptsize$1-x$} (X1);
    \draw (s4) edge[->] node[left] {\scriptsize$1-x$} (X2);
    \draw (s4) edge[->] node[right] {\scriptsize$x$} (X3);
    \draw (s2) edge[bend left, ->] node[left] {\scriptsize$y$} (s5);
    \draw (s2) edge[->] node[right] {\scriptsize$1-y$} (s6);
    \draw (s5) edge[bend left, ->] node[left] {\scriptsize$x$} (s2);
    \draw (s5) edge[->] node[left] {\scriptsize$1-x$} (X4);
    \draw (s6) edge[->] node[left] {\scriptsize$1-x$} (X5);
    \draw (s6) edge[->] node[right] {\scriptsize$x$} (X6);
    
    \node[draw=white, rectangle, fit=(current bounding box)] {};
\end{tikzpicture}
}
\subcaption{coins with unknown bias}
\label{fig:intro:ky:parametric}	
\end{subfigure}
\caption{Three variations of the Knuth-Yao die}
\label{fig:intro:psp}	
\end{figure}
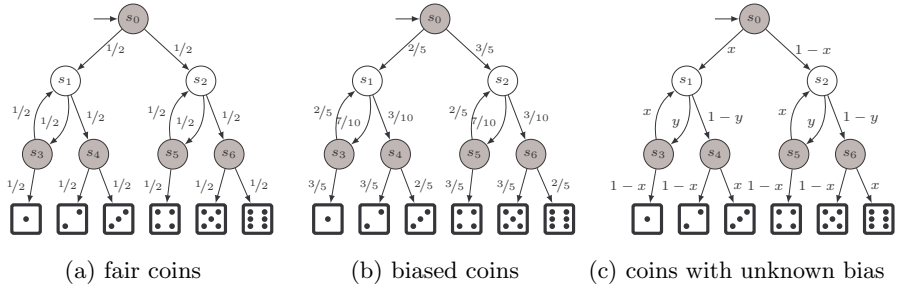

\noindent The parametric transition probability of going from state $s$ to $t$ is given by $\probdtmc(s, t)$.
Intuitively, instead of a concrete probability, a transition is equipped with a polynomial over the parameters $X$, e.g., $\probdtmc(s, t) = x^2{+}2{\cdot}y$.
A pMC with $X = \emptyset$ is a classical Markov chain (MC).
Applying an instantiation $v$ to a pMC $\pdtmc$ yields MC $\pdtmc[v]$ by replacing each transition $\probdtmc(s, t) = f \in \Q[X]$ in $\pdtmc$ by $f[v]$.
In general, a pMC defines an uncountable infinite family of MCs where each family member is obtained by a parameter instantiation. 
We assume every instantiation $v$ of $\pdtmc$ to yield well-defined MCs, i.e., $\probdtmc(s,\cdot)[v]$ is a probability distribution over the set $S$ of states; e.g., values $v(x)$ and $v(y)$ for $\probdtmc(s, t) = x^2{+}2y$ should be such that $0 \leq v(x)^2{+}2{\cdot}v(y) \leq 1$.
For a detailed treatment we refer to~\cite{DBLP:journals/corr/abs-1903-07993,sebi-diss-2020}.
An instantiation $v$ is \emph{graph-preserving} (for $\pdtmc$) if the topology of $\pdtmc$ is preserved, \ie, $\probdtmc(s,s') \neq 0$ implies $\probdtmc(s,s')[v] \neq 0$ for all $s, s' \in S$.
A region $R$ is graph-preserving if every $v \in R$ is graph preserving.

\begin{example}
\label{ex:psp}
Fig.~\ref{fig:kydie} depicts an MC with 13 states representing the Knuth-Yao algorithm~\cite{KY76} that mimics a six-sided die by repeatedly flipping a fair coin.
Fig.~\ref{fig:intro:ky:biased} shows a version that uses two coins (in the white and gray states respectively) with different but fixed biases.
Fig.~\ref{fig:intro:ky:parametric} provides a parametric version where the two coins have an unknown bias. 
The parameters $x$ and $y$ represent the probability to throw heads in the gray and white states, respectively.
The parameter space is $\{ (x,y) \mid 0 < x,y < 1 \}$.
The instantiation $v$ with $v(x) = \nicefrac{2}{5}$ and $v(y) = \nicefrac{7}{10}$ results in the MC in Fig.~\ref{fig:intro:ky:biased}.
The region of ``almost fair'' coins $x$ and $y$, say the coins that are fair up to a possible deviation of $\nicefrac{1}{10}$, is given by $\{ v \colon X \to \R \mid \nicefrac{9}{20} \leq v(x), v(y) \leq \nicefrac{11}{20} \}$.
\end{example}

\newpage

\section{Parameter Synthesis Problems}
\label{sec:problems}

\begin{wrapfigure}[12]{R}{4.1cm}
\vspace*{-1.0cm}
\hspace*{-0.2cm}
\includegraphics[scale=0.28]{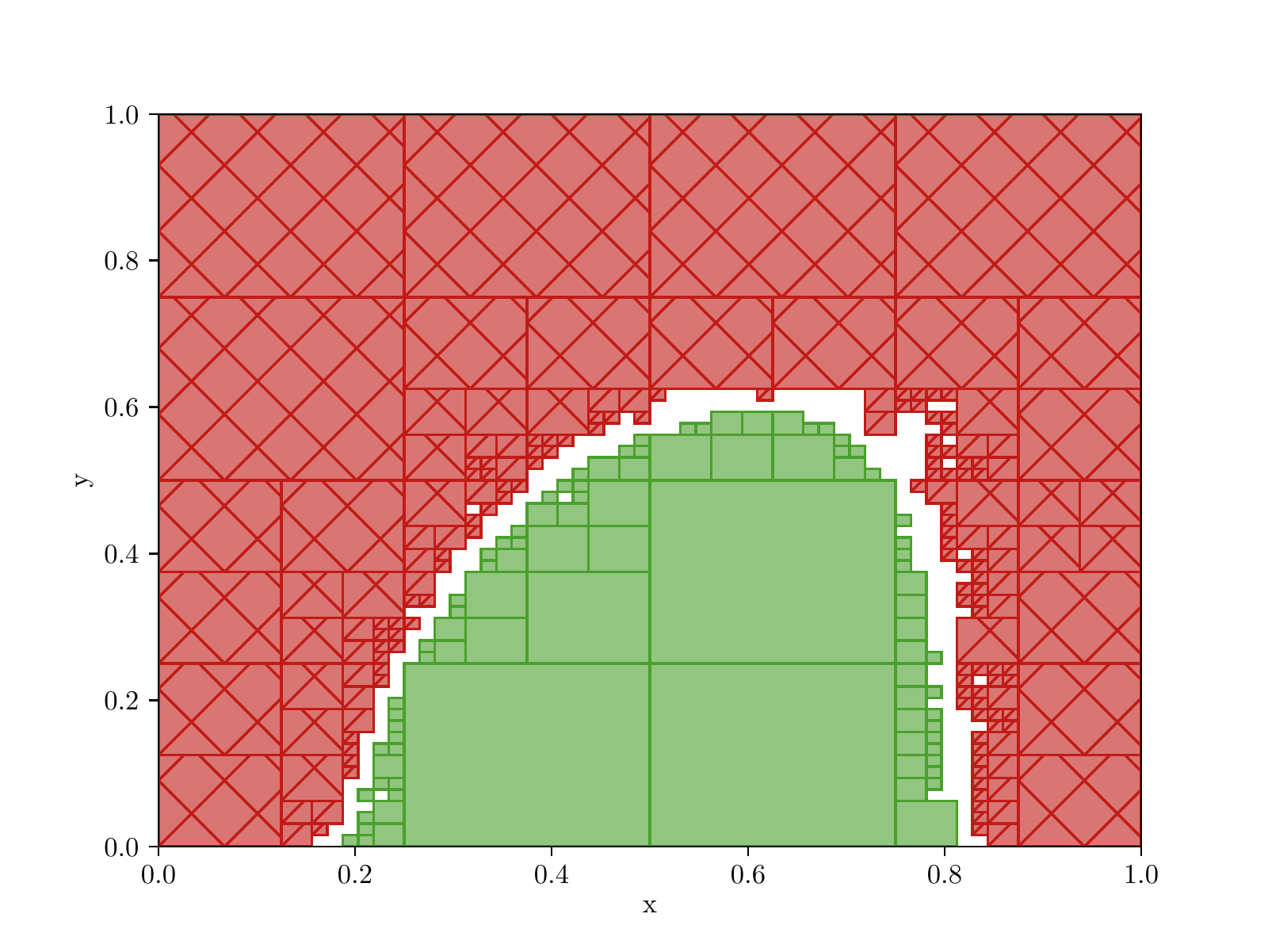}
\vspace*{-0.8cm}
\caption{Approximate partitioning to reach \drawdie[scale=0.7]{2} with probability $< \nicefrac{3}{20}$}
\label{parameter-partitioning-knuth-yao}
\end{wrapfigure}
A typical specification $\varphi$ requires to eventually reach a set of target states in pMC $\pdtmc$ with at least (or at most) a certain probability. 
A non-parametric MC $D$ satisfies $\varphi$, denoted $D \models \varphi$, if the reachability probability of the target states meets this threshold for its fixed transition probabilities.
In the presence of parameters, the validity of $\varphi$ is considered with respect to a given region $R$ of admissible parameter values.
Although this survey focuses on reachability probabilities this is not a severe restriction.
For instance, reachability probabilities suffice to determine the probability to satisfy an $\omega$-regular specification.
Such specifications contain LTL formulas and coincide with languages accepted by automata on infinite words such as non-deterministic B\"uchi automata.
The probability to satisfy $\omega$-regular specification $\varphi$ in a non-parametric MC $D$ equals the reachability probability of certain\footnote{The ones that are accepting according to the type (e.g., Muller, Rabin) of deterministic automaton to encode $\varphi$.}  bottom strongly connected components~\cite{Var85} in the synchronous product of $D$ with a deterministic automaton on infinite words corresponding to $\varphi$. 
Similar holds for the parametric setting.
Alternative specifications involve rewards (where $\pdtmc$ may have parametric rewards), bounded (aka: finite-horizon) reachability, etc.

Let pMC $\pdtmc$ over $X = \{x_1, \ldots, x_n\}$, specification $\varphi$, and region $R\subseteq \R^n$.
Let $v \in R$ stand for instantiation $v \colon X \to \R$ with $v(x_i) \in R{\upharpoonright}i$ for all $i$, and let:
$$
\pdtmc, R \models \varphi \quad \text{iff} \quad \left( \forall v \in R. \, \pdtmc[v] \models \varphi \right).
$$
We consider the following parameter synthesis problems:
\begin{itemize}
\item 
The \emph{feasibility problem} is to check whether:
$
\exists v \in R. \, \pdtmc[v] \models \varphi.
$  
Feasibility for $\varphi$ in $R$ is equivalent to not $\left( \pdtmc, R \models \neg\varphi \right)$.
\item The \emph{optimal} feasibility problem is to find an instantiation $v \in R$ that maximises (or dually minimises) the probability to satisfy $\varphi$.
\item 
The \emph{region verification problem} is to check whether: 
\[
\underbrace{\pdtmc, R \models \varphi}_{R \text{ is accepting}}
\quad \text{or} \quad
\underbrace{\pdtmc, R \models \neg\varphi}_{R \text{ is rejecting}}
\quad \text{or} \quad
\underbrace{\pdtmc, R \not\models \varphi  \ \land \ 
\pdtmc, R \not\models \neg\varphi}_{R \text{ is inconclusive}}.
\]
\item
The \emph{exact partitioning problem} is to partition (measurable) $R$ into $R_{+}$ and $R_{-}$ such that: 
\[
R_{+} = \{ \, v \in R \mid \pdtmc[v] \models \varphi \, \}
\quad \text{and} \quad
R_{-} = \{ \, v \in R \mid \pdtmc[v] \models \neg\varphi \, \}.
\]
Instead of checking whether a region $R$ is accepting or rejecting, the exact partitioning problem aims to find the largest subset of $R$ that is accepting (or, dually, rejecting).
As finding an exact partitioning is hard, often a relaxed version is considered.
\item The \emph{approximate partitioning problem} is to partition (measurable) $R$ into $R_{+}$, $R_{-}$, and $R_{?}$ such that for some given $0 \leq \eta \leq 1$:
\[
R_{+} \subseteq \{ \, v \in R \mid \pdtmc[v] \models \varphi \, \}
\quad \text{and} \quad
R_{-} \subseteq \{ \, v \in R \mid \pdtmc[v] \models \neg\varphi \, \},
\]
and $R_{?} = R \setminus (R_{+}  \, \cup \, R_{-})$ with $|| R_? || \leq (1{-}\eta) \cdot || R ||$ where $|| \cdot ||$ denotes the volume of $R$.
Here, $R_{?}$ is the fragment of $R$ that is inconclusive for $\varphi$. 
It is required that this fragment occupies at most a factor $1{-}\eta$ of the volume of $R$.
Stated differently, the accepting and rejecting region cover at least a factor $\eta$ of the region. 
Exact partitioning is obtained if $\eta = 1$.
\end{itemize}

\begin{example}\label{exa:synth-problems}
Consider the parametric die from Fig.~\ref{fig:intro:ky:parametric} and let $\varphi$ be the specification that the probability to reach state \drawdie[scale=0.7]{2} is at least $\nicefrac{3}{20}$.
There exists a feasible solution, e.g., $x = y = \nicefrac{1}{2}$.
Let region $R$ be all valuations with $\nicefrac{1}{10} \leq v(x) \leq \nicefrac{9}{10}$ and $\nicefrac{3}{4} \leq v(y)\leq \nicefrac{5}{6}$.
It follows that $R$ is accepting for $\neg \varphi$ (and thus rejecting $\varphi$) as for all $v \in R$ the probability to reach \drawdie[scale=0.7]{2} is at most $\nicefrac{3}{20}$.
The function:
\begin{equation}
\label{eqn:functionf}
f(x, y) = \frac{x \cdot (1-y) \cdot (1-x)}{1-x\cdot y}
\end{equation}
describes the probability to reach \drawdie[scale=0.7]{2}.
For $\neg\varphi$, $R_{+} = \{ v \mid f[v] < \nicefrac{3}{20} \}$ and $R_{-} = R \setminus R_{+}.$
Fig.~\ref{parameter-partitioning-knuth-yao} shows an approximation (for $\eta = 0.95$) of the function $f$ for $\neg\varphi$.
The set of rectangular accepting (green) regions indicate $R_{+}$ for $\neg \varphi$, whereas the red (hatched) area indicates $R_{-}$.
The white area indicates $R_{?}$.
\end{example}

The synthesis problems are complex as --- in contrast to interval MCs~\cite{DBLP:journals/rc/KozineU02} --- parameters may depend on each other and may occur at several transitions in a pMC.
These dependencies lead to trade-offs: increasing a parameter value may raise a reachability probability in one state but lower such probability in another state.
The next section addresses the theoretical complexity of the feasibility problem.

\section{Complexity of the Feasibility Problem}
\label{sec:complexity}

The ETR-SAT (Existential Theory of the Reals) decision problem consists of deciding whether a given existentially quantified formula $\exists x_1 \ldots \exists x_n. \, F(x_1, \ldots, x_n)$ holds, where $F$ is a Boolean combination of polynomial inequalities over the real-valued parameters $x_1$ through $x_n$.\footnote{In SAT modulo theories, ETR is referred to as the quantifier-free fragment of non-linear real arithmetic (QFNRA, for short).} 
The complexity class ETR contains all decision problems for which a polynomial many-to-one reduction to the ETR-SAT decision problem exists.
The class ETR contains NP and is contained in PSPACE~\cite{DBLP:conf/stoc/Canny88}.
Given that the Boolean structure of an ETR formula can be encoded into a polynomial, solving an ETR-SAT problem is equally hard as determining whether a polynomial over $x_1$ through $x_n$ has a real root.

\begin{figure}[h]
\centering
\vspace*{-0.4cm}
\includegraphics[scale=0.36]{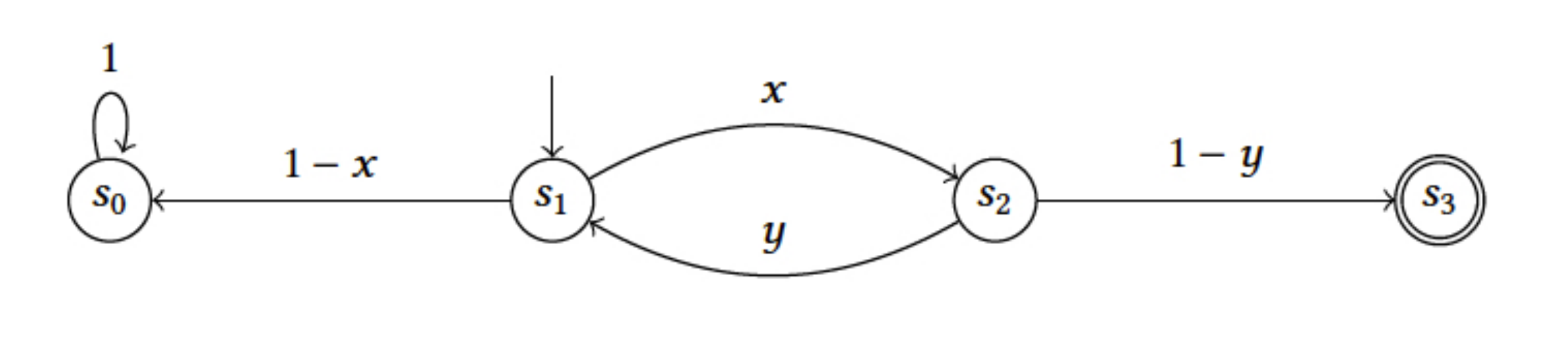}
\vspace*{-0.3cm}
\caption{A sample pMC for reachability probabilities}
\label{fig:example-pmc-reach}
\end{figure}

\paragraph{Encoding feasibility as ETR-SAT.}
The following example illustrates that the feasibility problem can be reduced to the ETR-SAT problem. 
\begin{example}\label{example:non-linear-equation-system}
Consider the pMC $\pdtmc$ in Fig.~\ref{fig:example-pmc-reach} with region $R = \{ v \mid 0 < v(x), v(y) < 1 \}$ and target state $s_3$. 
There exists a parameter instantiation such that the specification $\varphi$ holds, i.e., $s_3$ is reached with probability at least $\nicefrac{3}{4}$, whenever the following ETR-formula holds:
\[
\begin{array}{l}
\exists p_0, \ldots, p_3, x, y. \ 
\underbrace{0 < x < 1 \, \land \, 0 < y < 1}_{\footnotesize \mbox{region }R} \, \land \, 
\underbrace{p_1 \geq \nicefrac{3}{4}}_{\mbox{\footnotesize spec. } \varphi} \\[1ex]
\phantom{\exists p_0, \ldots, p_3, x, y.}
\ \wedge \ p_3 = 1 \ \wedge \ p_0 = 0 \\[1ex]
\phantom{\exists p_0, \ldots, p_3, x, y.} 
\ \wedge \ 
p_1 = x{\cdot} p_2 + \underbrace{(1{-}x){\cdot}p_0}_{= 0} 
\ \wedge \  
p_2 = y{\cdot}p_1 + \underbrace{(1{-}y){\cdot}p_3}_{= 1{-}y}.
\end{array}
\]
A variable $p_i$ is introduced for each state $s_i$ in $\pdtmc$ that intuitively represents the probability to eventually reach state $s_3$ from $s_i$.
Variables $x$ and $y$ represent the equally named parameters in $\pdtmc$.
The first line encodes the region $R$ and that $\varphi$ has to hold for state $s_1$.
The second line encodes that $s_3$ reaches itself almost surely, and that $s_3$ can almost surely not be reached from $s_0$.
The equations in the last line express the recursive equations for reachability probabilities.
They conform to the form:
\begin{align}
p_s \ = \ \sum_{s' \in S_{> 0}} \probdtmc(s,s') \cdot p_{s'} \quad
\text{for all } s \in S_{>0} \setminus G
\label{eq:etrencoding}
\end{align}
where $S_{>0} = \{ s_1, s_2, s_3 \}$ is the set of states that can reach a state in the set of target states $G = \{ s_3 \}$.
The size of the ETR-formula is linear in the size of $\pdtmc$. 
The set $S_{>0}$ can be computed by efficient graph analysis methods.
\end{example}

\paragraph{Encoding ETR-SAT as feasibility.}
Interestingly, this reduction also works (under mild conditions) in the reverse direction.
We illustrate by example how to obtain an acyclic pMC $\pdtmc_F$ from a given ETR-formula $F$ such that a target state can be reached with probability at least $\lambda$ in $\pdtmc_F$ for some variable instantiation $v$ if $F$ holds for $v$.
This reduction relies on important observations in~\cite{DBLP:journals/corr/Chonev17}.
The size of $\pdtmc_F$ is polynomial in the degree of $F$, the number of terms in $F$, and the maximal number of bits to encode the threshold $\lambda$ and $F$'s coefficients.

\begin{figure}[t]
\centering
\vspace*{-0.3cm}
\begin{tikzpicture}[initial text=,every node/.style={scale=1, font=\scriptsize}]
	\node[state, initial, scale=0.5] (s0) {};
	\node[state, right=of s0,yshift=0.8cm, scale=0.5] (a1) {};
	\node[state, right=of s0,yshift=0.3cm, scale=0.5] (b1) {};
	\node[state, right=of s0,yshift=-0.3cm, scale=0.5] (c1) {};
	\node[state, right=of s0,yshift=-0.8cm, scale=0.5] (d1) {};
	\node[state, right=of a1, scale=0.5] (a2) {};
	\node[state, right=of a2, scale=0.5] (a3) {};
	\node[state, right=of a3, scale=0.5, accepting] (a4) {};
	\node[state, right=of b1, scale=0.5] (b2) {};
	\node[state, right=of b2, scale=0.5,accepting] (b3) {};
	\node[state, right=of c1, scale=0.5,accepting] (c2) {};
	\node[state, right=of d1, scale=0.5,accepting] (d2) {};
	
	\draw[->] (s0) edge[bend left] node[pos=0.75,left,xshift=-1em] {$\nicefrac{2}{8}$} (a1);
	\draw[->] (s0) edge node[above,pos=0.8] {$\nicefrac{2}{8}$} (b1);
	\draw[->] (s0) edge node[above,pos=0.8] {$\nicefrac{2}{8}$} (c1);
	\draw[->] (s0) edge[bend right] node[pos=0.75,above] {$\nicefrac{1}{8}$} (d1);
	
	\draw[->] (a1) edge node[above] {$1-x$} (a2);
	\draw[->] (a2) edge node[above] {$x$} (a3);
	\draw[->] (a3) edge node[above] {$y$} (a4);
	
	\draw[->] (b1) edge node[above] {$1-x$} (b2);
	\draw[->] (b2) edge node[above] {$y$} (b3);
	
	\draw[->] (c1) edge node[above] {$1-y$} (c2);
	\draw[->] (d1) edge node[above] {$y$} (d2);

\end{tikzpicture}	
\caption{Sample construction for ETR-hardness for feasibility checking}
\label{fig:chonevtrick}
\end{figure}
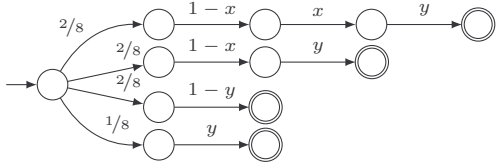

\begin{example}
Consider the ETR-formula $F = \exists x, y. \,  {-}2{\cdot}x^2{\cdot}y + y \geq 5$. 
We adapt this formula so as to remove the negative coefficients.
Simplification yields:	
\[ 2 \cdot (1{-}x) {\cdot} x {\cdot} y + 2 {\cdot} (1{-}x) {\cdot} y + 2{\cdot}(1{-}y) + y - 2 \ \geq \ 5.\]
By adding $2$ and dividing by $8$, the coefficients describe a probability distribution:	
\[   \nicefrac{2}{8} {\cdot} (1{-}x) {\cdot} x {\cdot} y + \nicefrac{2}{8} {\cdot} (1{-}x) {\cdot} y + \nicefrac{2}{8} {\cdot} (1{-}y) + \nicefrac{1}{8} {\cdot} y \ \geq \ \nicefrac{7}{8}.  \]
The left-hand side of this inequation induces the pMC in Fig.~\ref{fig:chonevtrick}, where the missing transitions all lead to a sink state (omitted in the figure).
The probability to reach a target state is at least $\nicefrac{7}{8}$ if the ETR-formula $F$ holds.
\end{example}

\begin{theorem}\label{th:etr-complete} \cite{DBLP:journals/jcss/JungesK0W21}
The feasibility problem for pMCs is ETR-complete.
\end{theorem}
\noindent
A few remarks are in order.
The reduction already applies to very simple classes of pMCs, e.g., acyclic ones, and pMCs that only contain expressions such as $x$ and $1{-}x$ on the transitions.
Our next remark concerns the complexity of related feasibility problems.
The above results apply to reachability objectives with non-strict conditions (i.e., either $\leq$ or $\geq$) on the probability thresholds.
For strict conditions, the feasibility problem is NP-hard.
The feasibility problem for qualitative reachability, i.e., thresholds 0 or 1, is NP-complete for most cases and in P for specific cases such as graph-preserving instantiations.
Our third remark concerns pMCs with non-deterministic choices.
The complexity class ETR is also relevant for synthesis problems on parametric MDPs; e.g., the feasibility problem ``does there exist a parameter valuation such that for some (or, dually, for all possible) resolution(s) of the non-determinism a reachability probability meets a non-strict threshold'' is ETR-complete too.
Full details can be found in~\cite{DBLP:journals/jcss/JungesK0W21}.
As a fourth remark, we point out that the above result holds for the setting with arbitrarily many parameters.
If the number of parameters is fixed, the feasibility problem for pMCs can be solved in polynomial time~\cite{DBLP:journals/corr/abs-1709-02093,DBLP:journals/iandc/BaierHHJKK20}.
The complexity of the fixed parameter case for parametric MDPs is still open.
Finally, we note that the above result indicates that solving the feasibility problem is difficult in general.
This motivates the search for heuristics and approximate techniques; such techniques are discussed in Section~\ref{sec:feasibility}.

\section{Algorithms}
\label{sec:algorithms}

This section describes the main underlying principles of algorithmic approaches to tackle the aforementioned synthesis problems: exact and approximate parameter space partitioning, region verification, and (optimal) feasibility checking.

\subsection{Exact Partitioning}

Exact partitioning amounts to computing the rational function $f$ as e.g., given in equation (\ref{eqn:functionf}).
This function is commonly referred to as the \emph{solution} function (for the reachability objective ``eventually \drawdie[scale=0.7]{2}''). 
Computing a closed-form  solution function has been one of the first problems considered for pMCs~\cite{Daws04} and has been subject to several practical efficiency improvements, see e.g., \cite{param_sttt,jansen-et-al-qest-2014,DBLP:journals/tse/FilieriTG16,DBLP:journals/corr/abs-1709-02093,DBLP:journals/iandc/BaierHHJKK20, DBLP:journals/corr/Chonev17,DBLP:conf/icse/FangCGA21}.
Solution functions map parameter values onto reachability probabilities.
For pMCs with polynomial transition functions as in this survey, solution functions are lower semi-continuous; for acyclic pMCs they are continuous.

\paragraph{Complexity.}
The functions can be obtained by successively eliminating the state variables from the characteristic equation system, e.g., the equations shown in  Example~\ref{example:non-linear-equation-system}.
This procedure is in fact a Gaussian elimination procedure on the parametric transition probability function.
Phrased differently, one solves a linear equation system with rational functions as coefficients.

\begin{example}
Consider again the pMC $\pdtmc$ in Fig.~\ref{fig:example-pmc-reach}.
The rational function for reaching $s_3$ is obtained by solving the non-linear equation system:
\[
p_0 = 0 \ \land \ p_1 = x{\cdot} p_2 \ \land \ p_2 = y{\cdot}p_1 + (1{-}y) \ \land \ p_3 = 1.
\]
The unique solution of this equation system yields:
\[
p_0 = 0 \ \land \ p_1 = \frac{x{\cdot}(1{-}y)}{1-x{\cdot}y} \ \land \ 
p_2 = \frac{1{-}y}{1-x{\cdot}y} \ \land \ p_3 = 1.
\]
The function of pMC $\pdtmc$ to reach $s_3$ from its initial state $s_1$ is thus $\frac{x{\cdot}(1{-}y)}{1-x{\cdot}y}$. 
\end{example}
The challenge is that elimination over the ring of polynomials results in large functions --- a problem that unfortunately cannot be avoided in general.

\begin{theorem}~\cite{DBLP:journals/corr/abs-1709-02093,DBLP:journals/iandc/BaierHHJKK20}
Computing the solution function (for reachability objectives) is exponential in the number of parameters, and polynomial in the number of states and the maximal degree of the polynomial transition probability functions.
\end{theorem}

\noindent
This result is a direct consequence of one-step fraction-free Gaussian elimination.
Standard Gaussian elimination does not suffice.
The point is that during triangulation, an elementary step in Gaussian elimination, the total degree of the involved factors is doubled in the worst case.
The growth of the fractions can be often limited by eliminating common factors, i.e., by keeping the numerator and denominator co-prime.
This, however, requires computing greatest common divisors which is expensive.\footnote{Data structures to avoid computing gcd's have been proposed in~\cite{jansen-et-al-qest-2014}.}
Fraction-free elimination avoids a fractional representation of the intermediate matrix entries altogether.
This does not avoid the exponential growth, but in combination with the one-step variant, it avoids gcd-computations.
This keeps the coefficients during the elimination process polynomially large.
Experiments~\cite{DBLP:journals/iandc/BaierHHJKK20} indicate that this seems beneficial for pMCs with a dense topology.
\begin{figure}[t]
\vspace*{-0.5cm}
\centering
\includegraphics[scale=0.5]{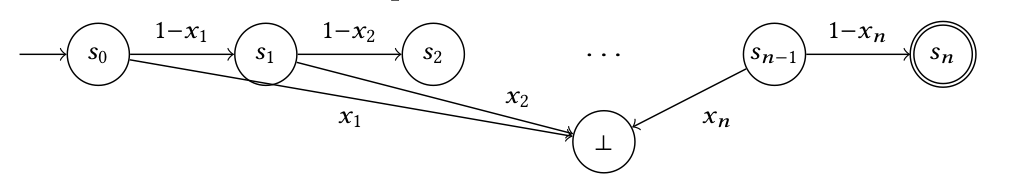}
\vspace*{-0.2cm}
\caption{Small multivariate pMC with an exponentially large solution function}
\label{fig:mall-pmc-with-large-sol-function}
\end{figure}

\begin{example}
The solution function for the family $\left( \pdtmc \right)_n$ of pMCs with $n \geq 2$ parameters, $n{+}2$ states and target state $s_n$ in Fig.~\ref{fig:mall-pmc-with-large-sol-function} is of the form:
\[
\prod_{i=1}^n (1-x_i) \ = \ \sum_{J \subseteq \{ 1, \ldots, n \}} -1^{|J|} \cdot \prod_{j \in J} x_j
\]
whose shortest sum-of-monomial representation has $2^n$ monomials.
\end{example}

\paragraph{State elimination.}
Rather than using (one-step fraction-free) Gaussian elimination, tools such as PARAM~\cite{param_sttt}, PRISM~\cite{KNP11} and Prophesy~\cite{DJJ+15} use state elimination on the pMC as originally proposed in~\cite{Daws04}.
This approach is analogous to computing regular expressions from non-deterministic finite automata (NFA)~\cite{HMU-Automata} and has also been applied in other contexts such as probabilistic workflow nets~\cite{DBLP:journals/pe/EsparzaHS17}.
The core idea behind state elimination is based on two operations: \\
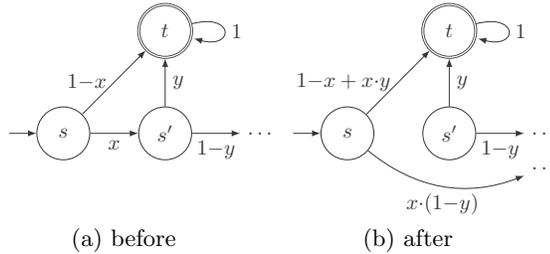
\begin{wrapfigure}[10]{l}{.63\textwidth}
\vspace*{-0.8cm}
\centering
\begin{subfigure}{0.3\textwidth}
\centering
\scalebox{0.8}{
\begin{tikzpicture}[baseline=(s0)]
		\node[state, , initial, initial text=] (s0) {$s$} ;
		\node[state, right=0.8cm of s0] (s1) {$s'$} ;
		\node[state, , above=0.8cm of s1, accepting] (s2) {$t$} ;
		\node[ right=0.8cm of s1] (s3) {$\hdots$} ;
		\node[below=0.3cm of s3] (s4) {\phantom{$\hdots$}} ;

		\draw[->] (s0) -- node[below] {$x$} (s1);
		\draw[->] (s1) -- node[right] {$y$} (s2);
		\draw[->] (s0) -- node[left] {$1{-}x$} (s2);
		\draw[->] (s1) -- node[below] {$1{-}y$} (s3);
		\draw[->] (s2) edge[loop right] node {$1$} (s2);
		\draw[->,white] (s0) edge[bend right] node[below] {\phantom{$x{\cdot}(1{-}y)$}} (s4);
	\end{tikzpicture}
}
\vspace*{-0.2cm}
\subcaption{before}
\label{fig:solutionfunction:shortcut}	
\end{subfigure}
\begin{subfigure}{0.3\textwidth}
\centering
\scalebox{0.8}{
\begin{tikzpicture}[baseline=(s0)]
		\node[state , initial, initial text=] (s0) {$s$} ;
		\node[state , right=0.8cm of s0] (s1) {$s'$} ;
		\node[state , above=0.8cm of s1, accepting] (s2) {$t$} ;
		\node[right=0.8cm of s1] (s3) {$\hdots$} ;
		\node[below=0.3cm of s3] (s4) {$\hdots$} ;
		
		\draw[->] (s1) -- node[right] {$y$} (s2);
		\draw[->] (s0) -- node[left] {$1{-}x + x{\cdot}y$} (s2);
		\draw[->] (s1) -- node[below] {$1{-}y$} (s3);
		\draw[->] (s2) edge[loop right] node {$1$} (s2);
		\draw[->] (s0) edge[bend right] node[below] {$x{\cdot}(1{-}y)$} (s4);
	\end{tikzpicture}
}
\vspace*{-0.2cm}
\subcaption{after}
\label{fig:solutionfunction:shortcut_applied}
\end{subfigure}
\vspace*{-0.2cm}
\caption{Adding short-cuts}
\end{wrapfigure}
(1) \emph{Adding short-cuts:}
Consider the pMC-fragment in Fig.~\ref{fig:solutionfunction:shortcut}. 
The reachability probabilities from any state to $t$ are as in Fig.~\ref{fig:solutionfunction:shortcut_applied} where the transitions from $s$ to $t$ via $s'$ are replaced by a short-cut from $s$ to $t$ and similar for all other direct successors of $s'$, thus bypassing $s'$. 
By successively creating short-cuts, eventually any path from the initial state to the target state consists of a single transition.

\begin{wrapfigure}[8]{l}{.63\textwidth}
\vspace*{-0.7cm}
\begin{subfigure}{0.3\textwidth}
\centering
\scalebox{0.8}{
\begin{tikzpicture}[baseline=(s1)]
	\node[state] (s1) {$s$};
	\node[state, right=1.2cm of s1] (s2) {$t$};
	\node[state, above=0.5cm of s2] (s3) {$t'$};
	\node[left=0.6cm of s1] (bbl) {};
	\node[right=0.9cm of s1] (bbr) {};
	\draw[->] (s1) edge[loop above] node[auto] {$z$} (s1);
	\draw[->] (s1) edge node[below] {$x$} (s2);
	\draw[->] (s1) edge node[auto] {$y$} (s3);
\end{tikzpicture}
}
\vspace*{-0.1cm}
\subcaption{before}
\label{fig:solutionfunction:with_loop}
\end{subfigure}
\begin{subfigure}{0.3\textwidth}
\centering
\scalebox{0.8}{
\begin{tikzpicture}[baseline=(s1)]
	\node[state] (s1) {$s$};
	\node[state, right=1.2cm of s1] (s2) {$t$};
	\node[state, above=0.5cm of s2] (s3) {$t'$};
	\node[left=0.6cm of s1] (bbl) {};
	\node[right=0.9cm of s1] (bbr) {};
	\draw[->] (s1) edge node[below] {$\nicefrac{x}{1{-}z}$} (s2);
	\draw[->] (s1) edge node[auto] {$\nicefrac{y}{1{-}z}$} (s3);
\end{tikzpicture}
}
\vspace*{-0.1cm}
\subcaption{after}
\label{fig:solutionfunction:without_loop}
\end{subfigure}
\vspace*{-0.2cm}
\caption{Self-loop removal}
\end{wrapfigure}
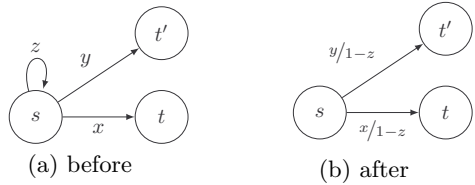
\noindent
(2) \emph{Eliminating self-loops:}
In order to bypass a state, it needs to be free of self-loops.
As the probability of staying forever in a non-absorbing state $s$ is zero, self-loops at $s$ can be eliminated by rescaling all outgoing transitions of $s$, cf.\ the change from Fig.~\ref{fig:solutionfunction:with_loop} to Fig.~\ref{fig:solutionfunction:without_loop}.

As for computing regular expressions from NFA, the elimination order of states is essential.
Computing an optimal order with respect to minimality of the result, however, is already NP-hard for acyclic NFA, see~\cite{DBLP:journals/fuin/Han13}.
Practical approaches resort to heuristics that are e.g., based on the pMC's topology, the in-degree of states, and so forth.
These heuristics can be static i.e., once fixed for the entire computation, or dynamic in which the ordering is re-considered during the computation depending on e.g., the complexity of transition probability functions.

\paragraph*{Experimental experiences.}
Extensive experiments are reported in~\cite[Section 10.2]{DBLP:journals/corr/abs-1903-07993}.
While solution functions can grow prohibitively large, medium-sized functions can still be computed.
E.g., for a variant of the NAND-example from the introduction with about 300K states and 400K  transitions, the solution function consisting of 4640 monomials of degree up to 200 could be computed in about 90 seconds. 
Experiments also reveal that the heuristic to determine the order of state elimination is essential and depends on the pMC at hand.
For most cases, state elimination (significantly) outperforms Gaussian elimination.
Recent experiments~\cite{DBLP:journals/corr/abs-2105-14371} indicate that the techniques described above significantly outperform similar techniques used for probabilistic graphical models; e.g., for Bayesian network \texttt{win95pts}\footnote{A network with a relatively high average Markov blanket, i.e., with relatively many causal dependencies between the random variables.} with 76 random variable and 200 parameters spread randomly over the graph, the solution function with about 7,5 million monomials of maximal degree 16 could be computed in about 400 seconds.
It is also beneficial to compute solution functions in a compositional way rather than on a monolithic pMC.
This has been advocated for each strongly connected component separately~\cite{jansen-et-al-qest-2014} and more recently for more liberal decompositions of the graphical structue of the pMC~\cite{DBLP:conf/icse/FangCGA21}.
Experiments in~\cite{DBLP:conf/icse/FangCGA21} indicate possible speed-ups of several orders of magnitude and more compact closed forms.

\subsection{Region Verification}
\label{sec:region}
\label{sec:parameter-lifting}

A direct approach to check whether $\pdtmc, R \models \varphi$ is to use the ETR-formulation:
one exploits a satisfiability modulo theory (SMT) checker for non-linear real arithmetic to check whether the conjunction of the ETR-formula $F$ as described in Section~\ref{sec:complexity} and a formula encoding the region $R$ is satisfiable.
The number of variables in the ETR-formula can be reduced significantly by pre-computing the solution function using the techniques described in the previous section.
The advantage of this approach is that it is exact and complete, but practically this SMT approach does not scale beyond a few parameters.

\begin{example}
For the pMC in Fig.~\ref{fig:example-pmc-reach}, this procedure amounts to check whether the ETR-formula in Example~\ref{example:non-linear-equation-system} holds, or equivalently whether, phrased using the solution function, $\frac{x{\cdot}(1{-}y)}{1-x{\cdot}y} \geq \nicefrac{3}{4}$ for some $0 < x,y < 1$.
\end{example}

\paragraph{A brief intermezzo: MDPs.}
An alternative is to exploit a \emph{dedicated abstraction technique}.
To that end, we require the notion of a (non-parametric) Markov decision process. 
Let $\Distr(S)$ be the set of discrete probability distributions over the set $S$ of states.
\begin{definition}[MDP]\label{def:mdp}
A Markov decision process (MDP) is a tuple $\mdp=(S,\sinit,\Act,P)$ with a finite set $S$ of \emph{states}, an \emph{initial state} $\sinit \in S$, a finite set $\Act$ of \emph{actions}, and a (partial) \emph{transition probability function} $P \colon S \times \Act \to \Distr(S)$.
\end{definition}
An MDP exhibits nondeterministic choices of actions at each state, where an action corresponds to a probability distribution over successor states, as in Markov chains.
So-called \emph{policies}, also referred to as schedulers or strategies, resolve this nondeterminism. 
For the type of specifications in this survey, it suffices to consider deterministic memoryless policies, specified by $\sigma \colon S\rightarrow \Act$.
Essentially, a policy $\sigma$ picks a unique action at each state of an MDP.
Model checking for MDPs employs techniques like value iteration or linear programming and aims to compute the maximal or minimal probability to satisfy a specification under any policy\cite{Put94,BK08}.
For surveys on MDP model checking, we refer to~\cite{DBLP:reference/mc/BaierAFK18,DBLP:series/lncs/BaierHK19}.

\paragraph{Parameter lifting.}
Coming back to parametric models, a viable and efficient method that works for a large class of pMCs is to transform a pMC $\pdtmc$ into an MDP whose minimal (maximal) reachability probability under-approximates (over-approximates) the reachability probability in $\pdtmc$ for all instantiations in a region.
The key steps of this so-called \emph{parameter lifting} approach~\cite{QDJJK16} are to:
\begin{itemize}
\item[(1)] relax the parameter dependencies (as inspired by \cite{BCDS13}), and 
\item[(2)] consider lower and upper bounds of parameters as worst cases.
\end{itemize}
Put in a nutshell, it reduces a region verification query on a pMC onto a verification query for an abstract probabilistic model, viz.\ an MDP representing an interval MC~\cite{DBLP:journals/ijar/Skulj09}.
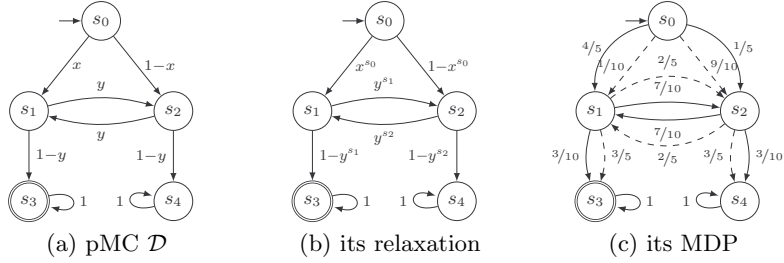
\begin{figure}[t]
\centering
\begin{subfigure}{0.3\textwidth}
\centering
\scalebox{0.8}{
\begin{tikzpicture}[scale=1, nodestyle/.style={draw,circle},baseline=(s0)]
    
    \node [nodestyle] (s0) at (0,0) {$s_0$};
    \node [] (leftdummy)  [on grid, left=1.2cm of s0] {};
    \node [] (rightdummy) [on grid, right=1.2cm of s0] {};
    \node [nodestyle] (s1) [on grid, below=\distforsone of leftdummy] {$s_1$};
    \node [nodestyle] (s2) [on grid, below=\distforsone of rightdummy] {$s_2$};
    \node [nodestyle, accepting] (s3) [on grid, below=\distsonesthree of s1] {$s_3$};
    \node [nodestyle] (s4) [on grid, below=\distsonesthree of s2] {$s_4$};
    
    \draw ($(s0)-(0.7,0)$) edge[->] (s0);
    \draw (s0) edge[->] node[right] {\scriptsize$x$} (s1);
    \draw (s0) edge[->] node[right] {\scriptsize$1{-}x$} (s2);

     \draw (s1) edge[bend left=15, ->] node[above] {\scriptsize$y$} (s2);
     \draw (s1) edge[->] node[right] {\scriptsize$1{-}y$} (s3);
    
     \draw (s2) edge[bend left=15, ->] node[below] {\scriptsize$y$} (s1);
     \draw (s2) edge[->] node[left] {\scriptsize$1{-}y$} (s4);
    
    \draw (s3) edge[loop right, ->] node[auto] {\scriptsize$1$} (s3);
    
    \draw (s4) edge[loop left, ->] node[auto] {\scriptsize$1$} (s3);

\end{tikzpicture}
}
\vspace*{-0.1cm}
\subcaption{pMC $\pdtmc$}
\label{fig:pmc-lifting-example}
\end{subfigure}  
\begin{subfigure}{0.3\textwidth}
\centering
\scalebox{0.8}{
\begin{tikzpicture}[scale=1, nodestyle/.style={draw,circle},baseline=(s0)]

    \node [nodestyle] (s0) at (0,0) {$s_0$};
    \node [] (leftdummy)  [on grid, left=1.2cm of s0] {};
    \node [] (rightdummy) [on grid, right=1.2cm of s0] {};
    \node [nodestyle] (s1) [on grid, below=\distforsone of leftdummy] {$s_1$};
    \node [nodestyle] (s2) [on grid, below=\distforsone of rightdummy] {$s_2$};
    \node [nodestyle, accepting] (s3) [on grid, below=\distsonesthree of s1] {$s_3$};
    \node [nodestyle] (s4) [on grid, below=\distsonesthree of s2] {$s_4$};
    
    \draw ($(s0)-(0.7,0)$) edge[->] (s0);
    \draw (s0) edge[->] node[right] {\scalebox{0.8}{$x^{s_0}$}} (s1);
    \draw (s0) edge[->] node[right] {\scalebox{0.8}{$1{-}x^{s_0}$}} (s2);

     \draw (s1) edge[bend left=15, ->] node[above] {\scalebox{0.8}{$y^{s_1}$}} (s2);
     \draw (s1) edge[->] node[right] {\scalebox{0.8}{$1{-}y^{s_1}$}} (s3);
    
     \draw (s2) edge[bend left=15, ->] node[below] {\scalebox{0.8}{$y^{s_2}$}} (s1);
     \draw (s2) edge[->] node[left] {\scalebox{0.8}{$1{-}y^{s_2}$}} (s4);
    
    \draw (s3) edge[loop right, ->] node[auto] {\scriptsize$1$} (s3);
    
    \draw (s4) edge[loop left, ->] node[auto] {\scriptsize$1$} (s3);
\end{tikzpicture}
}
\vspace*{-0.1cm}
\subcaption{its relaxation}
\label{fig:pmcrel:Dprime}
\end{subfigure}  
\begin{subfigure}{0.3\textwidth}
\centering
\scalebox{0.8}{
\begin{tikzpicture}[scale=1, nodestyle/.style={draw,circle},baseline=(s0)]
   
    \node [nodestyle] (s0) at (0,0) {$s_0$};
    \node [] (leftdummy)  [on grid, left=1.2cm of s0] {};
    \node [] (rightdummy) [on grid, right=1.2cm of s0] {};
    \node [nodestyle] (s1) [on grid, below=\distforsone of leftdummy] {$s_1$};
    \node [nodestyle] (s2) [on grid, below=\distforsone of rightdummy] {$s_2$};
    \node [nodestyle, accepting] (s3) [on grid, below=\distsonesthree  of s1] {$s_3$};
    \node [nodestyle] (s4) [on grid, below=\distsonesthree of s2] {$s_4$};
    
    \draw ($(s0)-(0.7,0)$) edge[->] (s0);
    \draw (s0) edge[->, bend left=0, dashed] node[left] {\scriptsize$\nicefrac{1}{10}$} (s1);
    \draw (s0) edge[->, bend right=0, dashed] node[right] {\scriptsize$\nicefrac{9}{10}$} (s2);
    \draw (s0) edge[->, bend right=40] node[left] {\scriptsize$\nicefrac{4}{5}$} (s1);
    \draw (s0) edge[->, bend left=40] node[right] {\scriptsize$\nicefrac{1}{5}$} (s2);

     \draw (s1) edge[bend left=40, ->, dashed] node[above] {\scriptsize$\nicefrac{2}{5}$} (s2);
     \draw (s1) edge[->, bend left=15, dashed] node[right] {\scriptsize$\nicefrac{3}{5}$} (s3);
     \draw (s1) edge[bend left=10, ->] node[above] {\scriptsize$\nicefrac{7}{10}$} (s2);
     \draw (s1) edge[->, bend right=15] node[left] {\scriptsize$\nicefrac{3}{10}$} (s3);
    
     \draw (s2) edge[bend left=40, ->, dashed] node[below] {\scriptsize$\nicefrac{2}{5}$} (s1);
     \draw (s2) edge[->, bend right=15, dashed] node[left] {\scriptsize$\nicefrac{3}{5}$} (s4);
     \draw (s2) edge[bend left=10, ->] node[below] {\scriptsize$\nicefrac{7}{10}$} (s1);
     \draw (s2) edge[->, bend left=15] node[right] {\scriptsize$\nicefrac{3}{10}$} (s4);
    
    \draw (s3) edge[loop right, ->] node[auto] {\scriptsize$1$} (s3);
    
    \draw (s4) edge[loop left, ->] node[auto] {\scriptsize$1$} (s3);
\end{tikzpicture}
}
\vspace*{-0.1cm}
\subcaption{its MDP}
\label{fig:approxmdp:M}
\end{subfigure}  
\caption{A two-phase procedure to transform a pMC into an MDP}
  \label{fig:parameter-lifting}
\end{figure}

\begin{example}
Consider the pMC $\pdtmc$ in Fig.~\ref{fig:pmc-lifting-example} and let region $R = [\nicefrac{1}{10}, \nicefrac{4}{5}] \times [\nicefrac{2}{5}, \nicefrac{7}{10}]$\footnote{More precisely, $R = \{ v \mid v(x) \in [\nicefrac{1}{10}, \nicefrac{4}{5}], v(y) \in [\nicefrac{2}{5}, \nicefrac{7}{10}]\}$.}.
The parameter $y$ occurs at transitions emanating from two states.
This leads to the following trade-off.
To maximise reaching the state $s_3$ from $s_1$, the value of $y$ should be low.
However, from state $s_2$, $y$'s value should be large so as to avoid reaching $s_4$ from which $s_3$ is unreachable.
The optimal $y$-value depends on the probability to visit $s_1$ and $s_2$ and thus directly on $x$'s value.

(1) The first conceptual step in parameter lifting is to remove parameter dependencies: 
We make all parameters occurrences unique for each state.
Parameter $y$ at state $s_1$ becomes $y^{s_1}$ and at $s_2$ becomes $y^{s_2}$, etc.
Fig.~\ref{fig:pmcrel:Dprime} shows the resulting pMC $\mbox{\sf rel}(\pdtmc)$.
The corresponding 3-dimensional relaxed version of $R$ is: 
\[
\mbox{\sf rel}(R) \ = \ 
\underbrace{[\nicefrac{1}{10}, \nicefrac{4}{5}]}_{\text{range of } x^{s_0}} \times 
\underbrace{[\nicefrac{2}{5}, \nicefrac{7}{10}]}_{\text{range of } y^{s_1}} \times
\underbrace{[\nicefrac{2}{5}, \nicefrac{7}{10}]}_{\text{range of } y^{s_2}}. 
\]
On instantiating the pMC $\mbox{\sf rel}(\pdtmc)$, the ``local'' copies of $y$ get the same value.
Thus, e.g., $(v(x), v(y)) = (\nicefrac{4}{5}, \nicefrac{3}{5}) \in R$ becomes $(\rel{v}(x^{s_0}), \rel{v}(y^{s_1}), \rel{v}(y^{s_2})) = (\nicefrac{4}{5}, \nicefrac{3}{5}, \nicefrac{3}{5}) \in \mbox{\sf rel}(R)$.
The region $\rel{R}$ contains spurious instantiations, e.g., $(\nicefrac{4}{5}, \nicefrac{1}{2}, \nicefrac{3}{5})$, where the two copies of $y$ have distinct values.
These instantiations do not have a counterpart in $R$.
\emph{Removing parameter dependencies is thus an over-approximation}.

(2) The second conceptual step creates the MDP in Fig.~\ref{fig:approxmdp:M}.
It has the same state space as the pMC $\mbox{\sf rel}(\pdtmc)$.
States $s_0$ through $s_2$ are equipped with actions for upper (lower) bounds in the region $\mbox{\sf rel}(R)$, indicated by solid (dashed) lines.  
For instance, state $s_2$ has a solid (dashed) transition with probability $\nicefrac{7}{10}$ ($\nicefrac 2 5$) to $s_1$.
These transitions represent the largest (smallest) probability to go from $s_2$ to $s_1$ in $\mbox{\sf rel}(R)$.
As each value between these two extreme values can be selected, the resulting MDP is in fact an interval MC~\cite{DBLP:journals/ijar/Skulj09}.
Choices in $s_3$ and $s_4$ are unique, as the outgoing transitions in $\mbox{\sf rel}(\pdtmc)$ are constant.

How is an analysis of the resulting MDP related to checking whether $\pdtmc, R \models \varphi$?
Assume that $\varphi$ requires to reach state $s_3$ with at most probability $\nicefrac{4}{5}$, i.e., $\nicefrac{48}{60}$.
Verifying the resulting MDP using standard probabilistic model checking (PMC) techniques~\cite{katoen2016probabilistic,DBLP:reference/mc/BaierAFK18,DBLP:series/lncs/BaierHK19} yields a maximal probability to reach $s_3$ of $\nicefrac{47}{60}$.
This is achieved by a policy that selects solid transitions in $s_0$ and $s_2$ and dashed transitions in $s_1$.
As $\nicefrac{47}{60} \leq \nicefrac{48}{60}$, and the MDP over-approximates pMC $\pdtmc$, it follows that $\pdtmc, R \models \varphi$.
If $\varphi$ would impose a lower bound (rather than an upper bound) on the reachability objective, the minimal probability to reach $s_3$ in the MDP is considered.
\end{example}
 
\paragraph{Correctness.}
For the following result, we use the notion of the Markov chain $\mdp^\sigma$ that is \emph{induced} by an MDP $\mdp$ and a policy $\sigma$.


\begin{theorem} \cite{QDJJK16}
Let $\pdtmc$ be a locally monotone pMC, $R$ a rectangular and closed region\footnote{A rectangular and closed region is of the form $\prod_i [\ell_i, u_i]$ where the value of parameter $x_i$ lies in the closed interval $[\ell_i,u_i]$.}, $\varphi$ a reachability specification, and $\mdp$ the resulting MDP from lifting $\pdtmc$ and $R$.
Then:
\begin{enumerate}
\item (for all policies $\sigma$ on $\mdp$: $\mdp^\sigma \models \varphi$) $\quad \quad \text{implies} \quad$
$\pdtmc, R \models \varphi$, and
\item (for all policies $\sigma$ on $\mdp$: $\mdp^\sigma \models \neg\varphi$) $\quad \phantom{i} \text{implies} \quad$
$\pdtmc, R \models \neg\varphi,$
\end{enumerate}
where $\sigma$ ranges over all possible
policies for the MDP $\mdp$.
\end{theorem}

Note that the (graph-preserving) region $R$ is required to be rectangular and closed, and that pMC $\pdtmc$ is locally monotone, i.e., the transition probability functions in $\pdtmc$ are monotonic.
Examples of such functions are e.g., multi-linear polynomials.
The first constraint makes the bounds of the parameters independent of other parameter instantiations and ensures the maximum (and minimum) over the region $R$ to exist.
Due to the local monotonicity and the absence of parameter dependencies after relaxation, maximal reachability probabilities are easy to determine: it suffices to consider instantiations that set the value of each parameter to either the lowest or highest possible value.
These two constraints on $\pdtmc$ and $R$ are quite mild; in fact, almost all examples from the literature comply to these constraints.

\paragraph{Region refinement.}
If the over-approximation $\mbox{\sf rel}(R)$ of region $R$ is too coarse for a conclusive answer, $R$ can be refined into smaller regions.
Intuitively, by excluding more potential parameter values, the actual choice of the parameter value has lesser impact on reachability probabilities.
The smaller the region, the smaller the over-approximation: the optimal instantiation on the pMC $\pdtmc$ is over-approximated by some policy on the MDP $\mdp$. 
The approximation error originates from choices where an optimal policy on $\mdp$ chooses actions $v_1$ and $v_2$ at states $s_1$ and $s_2$, respectively, with $v_1(x_i^{s_1}) \neq v_2(x_i^{s_2})$ for some parameter $x_i$, and therefore intuitively disagree on its value.
The probability mass that is affected by these choices decreases if the region is smaller.

\paragraph{Experimental results.}
Experiments~\cite{QDJJK16,DBLP:journals/corr/abs-1903-07993} show that parameter lifting is an excellent approach for region verification if the influence of the parameter dependencies is minor, or if the number of parameters is relatively low (at most 15, say).
For more parameters, the refinement in exponentially many regions often yields a prohibitive number of sub-regions that have to be analysed.

The beauty of parameter lifting is threefold: (1) it is conceptually simple, (2) its principle is applicable to other models such as parametric MDPs (which are then transformed into two-player stochastic games), and (3) it reduces a verification problem on parametric probabilistic models to a model-checking query on non-parametric probabilistic models for which efficient algorithms exist.

\subsection{Approximate Parameter Space Partitioning}

The approximate parameter space partitioning problem requests to partition region $R$ into accepting and rejecting fragments together with an unknown fragment which covers at most fraction $1{-}\eta$ of $R$'s volume.
This section presents an iterative approach to this problem that builds on top of region verification discussed just above. 
The basic idea of this approach was first presented in~\cite{DJJ+15} and has substantially been refined and extended in \cite[Ch.\ 8]{sebi-diss-2020}.

\paragraph{The basic idea.}
Approximate parameter space partitioning can be viewed as a \emph{counter-example guided abstraction} refinement (CEGAR)-like~\cite{CGJLV00} approach to successively divide the parameter space into accepting and rejecting regions.
The basic idea is to compute a sequence $\left( R^i_{+} \right)_{i \in \N}$ of simple (read: rectangular) accepting regions that successively extend each other.
Similarly, an increasing sequence $\left( R^i_{-} \right)_{i \in \N}$ of simple rejecting regions is computed.
At the $i$-th iteration, $R^i = R^i_{+} \cup R^i_{-}$ is the covered fragment of the parameter space.
The iterative approach halts when $R^i$ covers at least $100{\cdot}\eta$\% of the entire region $R$.
Termination is guaranteed: in the limit a solution to the exact synthesis problem is obtained as $\lim_{i \rightarrow \infty} R_{+}^i = R_{+}$ and $\lim_{i \rightarrow \infty} R_{-}^i = R_{-}$. 

Issues are e.g., how to find good region candidates, how to refine a region, and how to effectively use diagnostic feedback if a region is found to be inconclusive.

\begin{example}[A naive procedure]
\label{ex:naiverefinement}
Consider the parametric die from Example~\ref{ex:psp}. 
Suppose we want to synthesise the partitioning in Fig.~\ref{parameter-partitioning-knuth-yao} on page~\pageref{parameter-partitioning-knuth-yao}. 
We start by verifying the full parameter space $R$ against $\varphi$.
The verifier returns \texttt{false}, as $R$ is not accepting. 
Since $R$ might be rejecting, we invoke the verifier with $R$ and $\neg\varphi$, yielding \texttt{false} too.
Thus, the full parameter space $R$ is inconclusive.
Let us now split $R$ into four equally-sized regions, all of which turn out to be inconclusive. 
Only after splitting again by the same principle, the first accepting and rejecting regions are found.
After various iterations, this procedure leads to the partitioning in Fig.~\ref{parameter-partitioning-knuth-yao}.
\end{example}

\begin{figure}[t]
\centering
\vspace*{-0.7cm}
\begin{subfigure}{0.32\textwidth}
\centering
\includegraphics[scale=0.26]{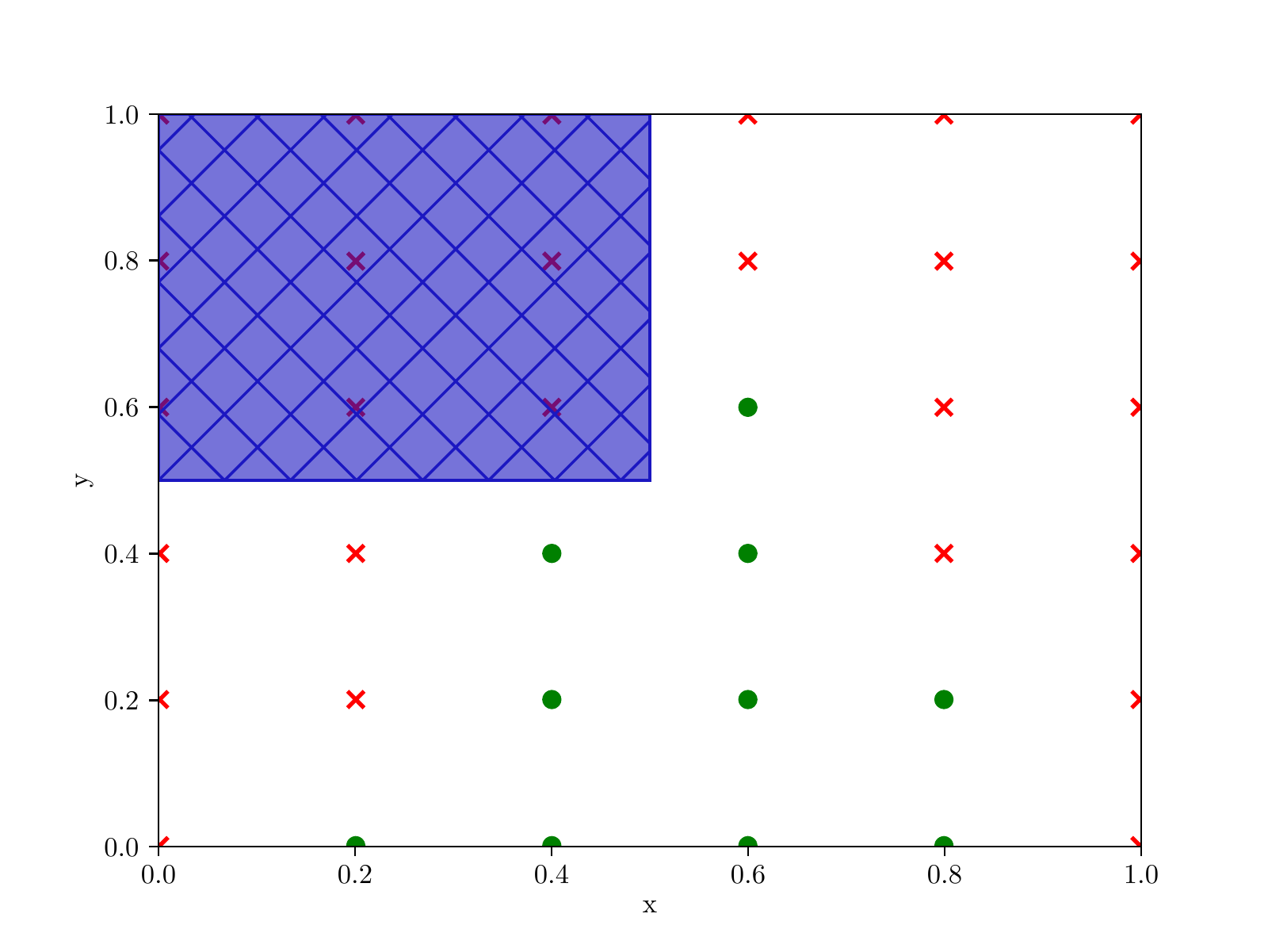}
\vspace*{-0.4cm}
\subcaption{(Uniform) sampling}
\label{fig:pspex:1}
\end{subfigure} 
\begin{subfigure}{0.32\textwidth}
\centering
\includegraphics[scale=0.26]{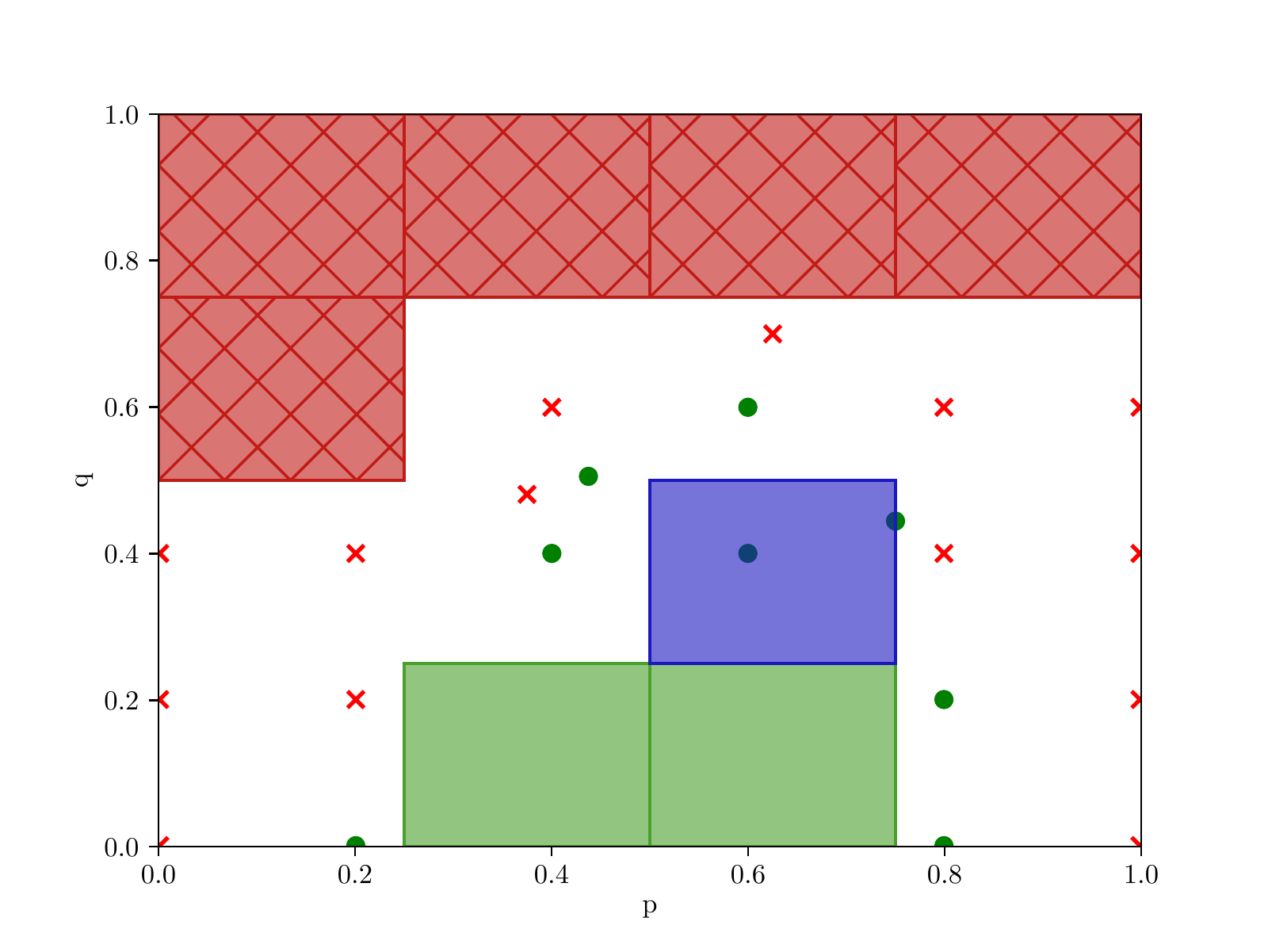}
\vspace*{-0.4cm}
\subcaption{Generating candidates}
\label{fig:pspex:2}
\end{subfigure} 
\begin{subfigure}{0.32\textwidth}
\centering
\includegraphics[scale=0.26]{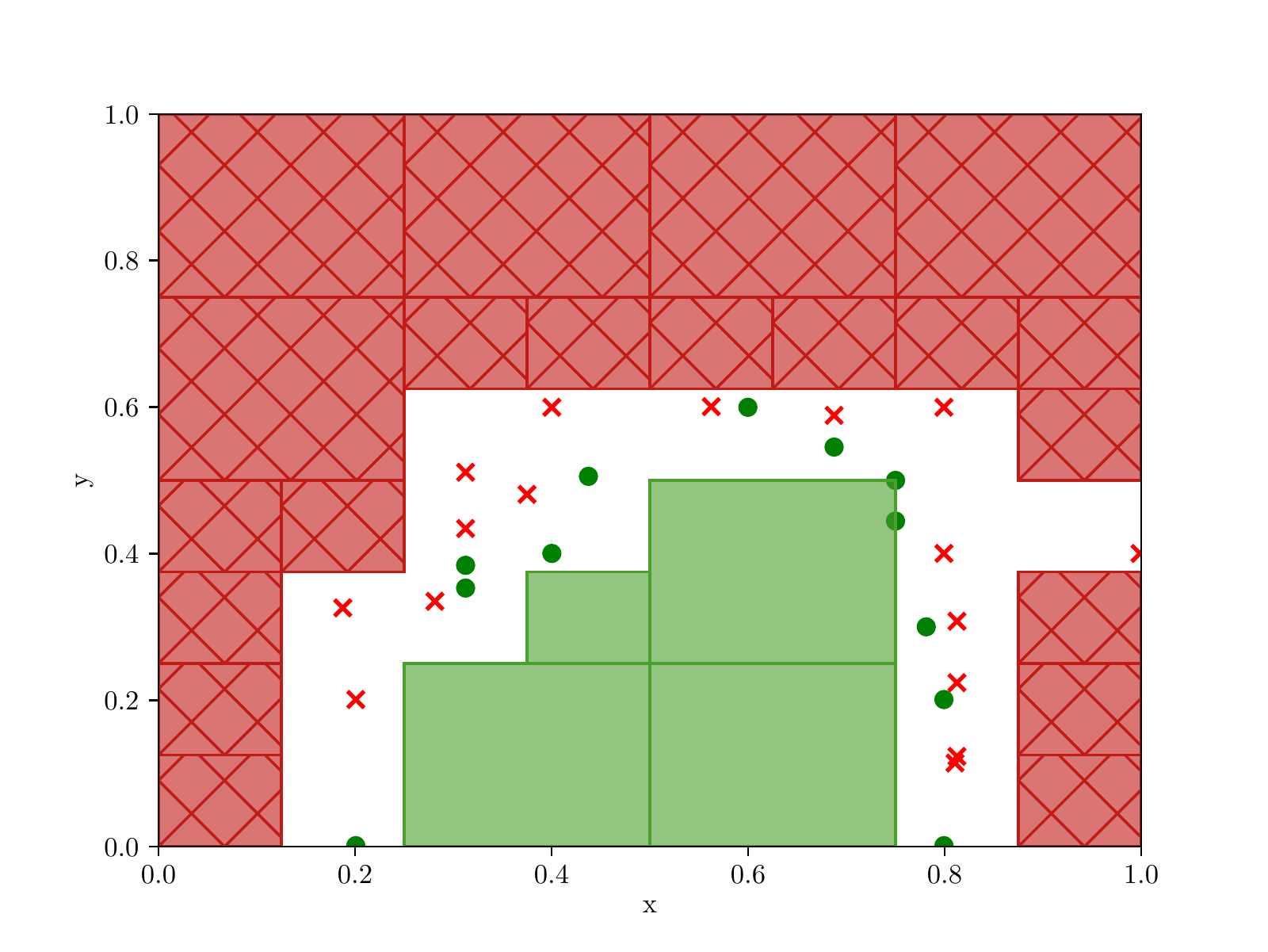}
\vspace*{-0.4cm}
\subcaption{Preliminary result}
\label{fig:pspex:3}
\end{subfigure} 
\vspace*{-0.2cm}
\caption{Parameter space partitioning in progress (using \prophesy~\cite{DJJ+15}.)}
\end{figure}

\paragraph{Finding region candidates.}
Rather than starting with the entire (typically large and inconclusive) initial region $R$, a better strategy is to first do some \emph{sampling}. 
Here, sampling means identifying an instantiation $v \in R$ and verifying using an off-the-shelf probabilistic model checker for MCs whether $\pdtmc[v] \models \varphi$.
One can e.g., start by uniformly sampling the region $R$, see Fig.~\ref{fig:pspex:1}.
A red cross at point $v = (v(x), v(y))$ indicates that $\pdtmc[v] \models \neg\varphi$, while a green dot indicates $\pdtmc[v] \models \varphi$.
The sampling results can be used to steer the selection of a candidate region that is consistent with all samples contained in it; e.g., the blue rectangle in Fig.~\ref{fig:pspex:1} is a candidate region for the hypothesis $\neg\varphi$ (indicated by the hatching) as it only contains red crosses.
\paragraph{How to split regions?}
A simple strategy is to split regions that sampling or region verification revealed to be inconsistent.

\begin{example}
After several iterations, the iterative refinement proceeded to the situation in Fig.~\ref{fig:pspex:2}.
The blue candidate region in Fig.~\ref{fig:pspex:1} turned out to be not rejecting; the region verifier provided the counterexample $(p,q) = (0.45, 0.52)$.
Further iterations with smaller regions had some success, but some additional samples (the dots and circles that are off-grid points) were obtained as counterexamples. 
The blue region in Fig.~\ref{fig:pspex:2} is a candidate for $\varphi$.
Fig.~\ref{fig:pspex:3} shows a further snapshot indicating that this blue candidate region is indeed accepting. 
The white box on the right border has been checked but its verification timed out without any counterexample.
\end{example}
Splitting of regions based on the available samples can be done in various ways, e.g., by splitting regions into equally-sized regions (as used in the above example), or by attempting to gradually obtain larger candidates.
The rationale of the former is to obtain small regions with small bounds whereas the latter is based on the rationale to quickly cover a vast fragment of the parameter space.
For the construction of region candidates, a good practical strategy is to split the initial regions according to a heuristic until none of the regions is inconclusive. 
Candidate regions are sorted in descending size.
Regions are preferred where verification seems less costly: candidate regions that are supposed to be accepting and are further away from samples or regions that are rejecting are preferred over those regions which have rejecting samples or are close to rejecting regions.
\begin{figure}[h]
\begin{subfigure}{0.48\textwidth}
\centering
\includegraphics[scale=0.66]{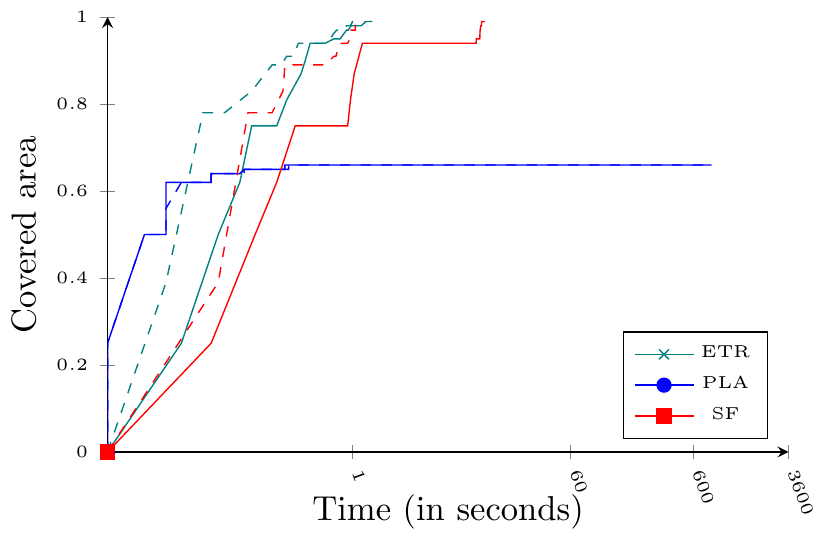}
\vspace*{-0.2cm}
\subcaption{Herman's model}
\label{fig:results_covered_area_herman}
\end{subfigure} 
\begin{subfigure}{0.48\textwidth}
\centering
\includegraphics[scale=0.66]{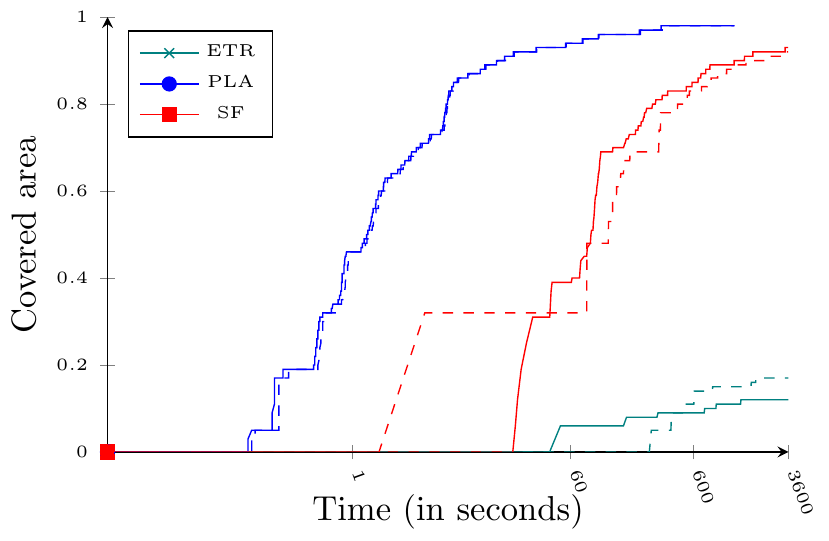}
\vspace*{-0.2cm}
\subcaption{NAND model}
\label{fig:results_covered_area_nand_03}
\end{subfigure} 
\vspace*{-0.1cm}
\caption{Covered parameter space during partitioning for two benchmarks}
\label{fig:results_covered_areas}
\end{figure}

\paragraph{Experimental experiences.}
Fig.~\ref{fig:results_covered_area_herman} depicts results for parameter space partitioning on Herman's randomised self-stabilisation algorithm~\cite{DBLP:journals/ipl/Herman90} for a distributed system with five processes and threshold $\lambda=5$.
This model has one parameter, 33 states and 276 transitions.
The plot depicts the covered parameter space for three techniques with both quads (straight lines) and rectangles (dashed lines) as region representations.
A point $(x,y)$ indicates that $y$ percent of the parameter space could be covered within $x$ seconds.
The three techniques are parameter lifting (PLA), the ETR-encoding, and the solution function (SF) encoded as ETR-formula.
Recall that the ETR-encoding contains a variable for each state and each parameter, whereas the SF only contains a variable for each parameter, cf.\ equation (\ref{eqn:functionf}).

\begin{wrapfigure}[15]{r}{0.5\textwidth}
\vspace*{-0.4cm}
\centering
\includegraphics[scale=0.45]{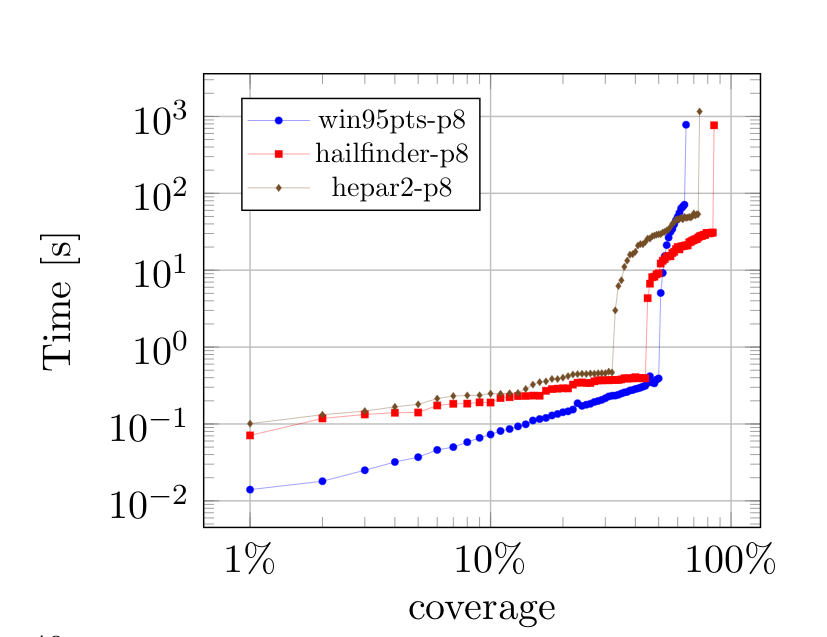}
\vspace{-0.6cm}
\caption{Approximate parameter partitioning on three Bayesian networks with each eight parameters}
\label{fig:plot-pla-bayes}
\end{wrapfigure}
For Herman's distributed algorithm, SMT-based techniques perform better than PLA.
PLA was able to cover 64\% of the parameter space within milliseconds but only 2\% more space was covered in the remaining hour.
SMT was able to cover at least 99\% of the parameter space within 15 seconds.
Moreover, rectangles cover the parameter space faster than quads.
The results on the NAND model~\cite{DBLP:journals/pieee/HaselmanH10} are a bit different, see Fig.~\ref{fig:results_covered_area_nand_03}.
This model has two parameters, 178 states and 243 transitions.
Here, PLA outperforms SMT.
Rectangles and quads perform on par, although for smaller thresholds their performance does differ. 
Applications on various benchmarks reveals that parameter lifting is almost always superior but not on all benchmarks and all (sub-)regions.
There can be substantial differences between the heuristics for candidate generation, especially in settings where single region verification calls are expensive.
To indicate the scalability of the approach, we report on some recent experiments~\cite{salmanikatoen2022} on parametric Bayesian networks. Figure~\ref{fig:plot-pla-bayes} shows that 80-90\% of the 8-dimensional parameter space for three parametric Bayesian networks (\texttt{win95pts}, \texttt{hailfinder}, and \texttt{hepar-2}) can be covered in about 100-1,000 seconds.

\subsection{Feasibility Checking}
\label{sec:feasibility}

We now return to the ETR-complete feasibility problem of before and describe how techniques from mathematical optimisation can be exploited to obtain a scalable approach.

\begin{figure}[t]
    \centering
    \vspace*{-1.5cm}
    \rotatebox{270}{\includegraphics[scale=0.36]{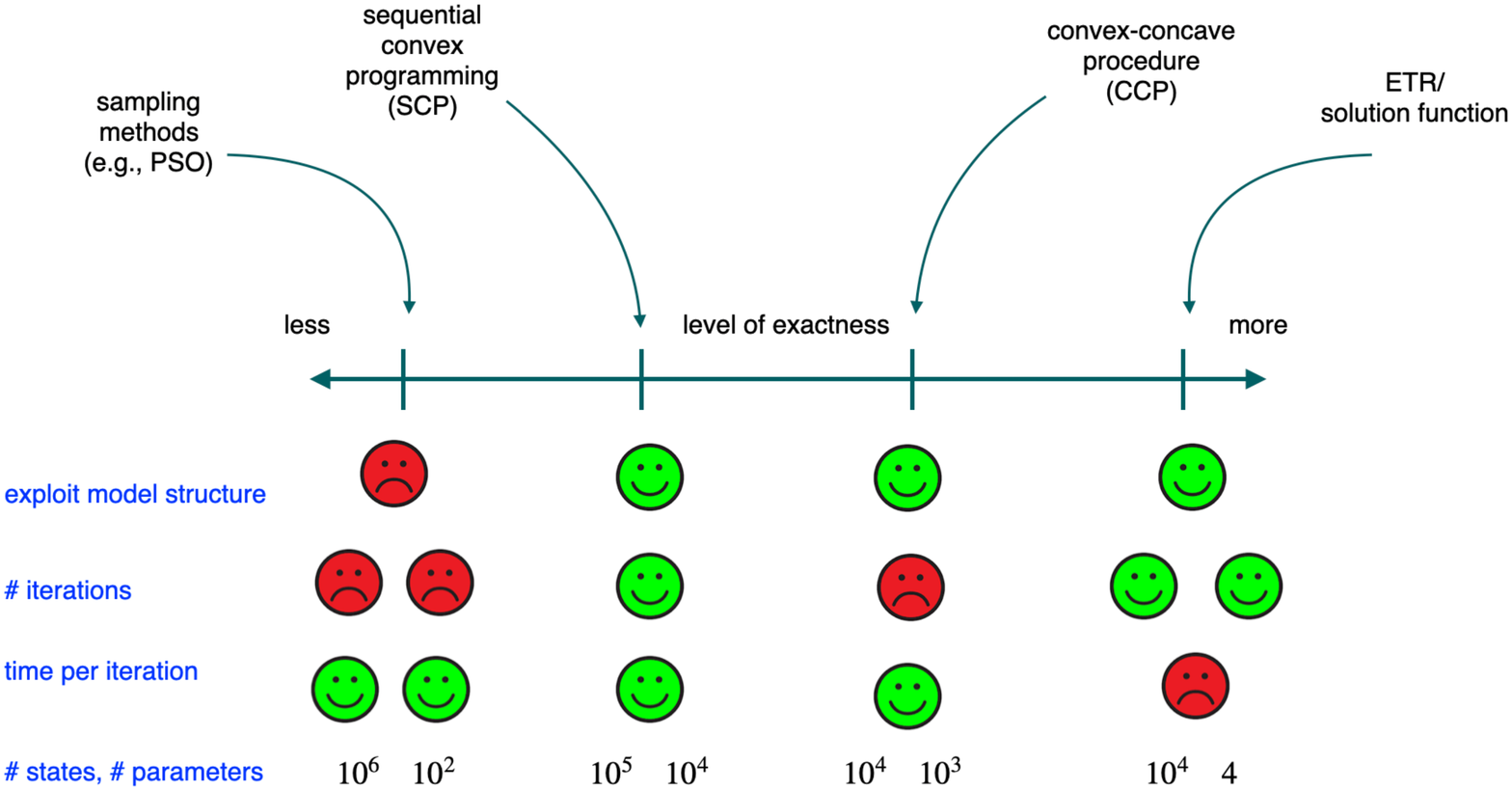}}
    \vspace*{-1.0cm}
    \caption{Comparing various techniques for feasibility for different levels of precision}
    \label{fig:feasibility-spectrum}
\end{figure}

\paragraph{Two ``extreme'' approaches: sampling and ETR-solving.}
To find a satisfiable parameter instantiation it suffices to guess an instantiation $v$ such that $\pdtmc[v] \models \varphi$.
Checking whether $\pdtmc[v] \models \varphi$ can be done using PMC techniques~\cite{katoen2016probabilistic,DBLP:reference/mc/BaierAFK18} or, alternatively, by instantiating the solution function efficiently~\cite{DBLP:conf/atva/GainerHS18}.
This insight has led to an adaptation of sampling-based techniques to the parameter synthesis problem, most prominently particle swarm optimisation (PSO)~\cite{chen2013model}. 
These techniques can handle millions of states but their efficiency is limited for models with more than roughly ten parameters, see the left part of Fig.~\ref{fig:feasibility-spectrum}.
Rather than pure sampling-based techniques such as PSO, one can pursue an exact (but costly) approach by computing the ETR-formula and checking whether it is satisfiable.
This is exact, but costly as indicated on the right part of Fig.~\ref{fig:feasibility-spectrum}.

\paragraph{A middle ground: mathematical optimisation.}
An approach that is suitable compromise between sampling and exact solution techniques is to exploit techniques from \emph{mathematical optimisation}.
The basic idea is to guess a parameter instantiation $v$, and then optimise around this $v$ in the parameter space so as to find a point that improves the reachability probability (or whatever specification we are interested in) in the induced MC. 
This is repeated until an accepting instantiation is found. 
Rather than optimising on the original ETR formula, the formula is approximated such that every iteration step is computationally tractable. 
This idea can be realised using different ideas. 
These ideas share that they exploit the model structure of the pMC $\pdtmc$, and need fewer iterations than sampling-based methods.
Differences occur due to the choice of approximation, e.g., the problem can be convexified~\cite{DBLP:conf/atva/CubuktepeJJKT18} or linearised~\cite{DBLP:journals/corr/abs-2107-00108}.
Although we consider pMCs, this technique is also applicable to parametric MDPs.

\begin{figure}[t]
\centering
\includegraphics[scale=0.6]{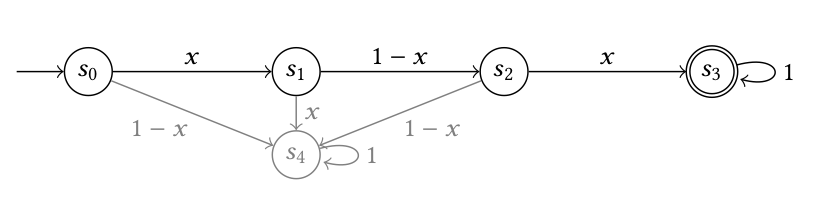}
\vspace*{-0.3cm}
\caption{A sample pMC to illustrate the use of mathematical optimisation}
\label{fig:example-pmc-qcqp}
\vspace*{-0.5cm}
\end{figure}

\paragraph{Feasibility as an optimisation problem.}
The first step is to formulate the feasibility problem as an optimisation problem.  
Assuming that all transition probabilities are linear in the parameters, the ETR-formula can be represented as a  \emph{quadratically-constrained quadratic program} (QCQP~\cite{BV2004}), an optimisation problem with a quadratic objective that is to be achieved under a set of quadratic constraints\footnote{For readers familiar with the LP encoding for reachability objectives for MDPs, this QCQP is the straightforward representation of this LP where the transition probabilities (parameters) are multiplied with the state-variables.}.
We show this for a key part of the ETR encoding:
\begin{align}
\begin{split}
\label{eq:etr-encoding} 
& \text{minimise } p_{\sinit} \text{ s.t.}\quad \\
& p_s \ \geq \ \sum_{s' \in S_{> 0}} 
\underbrace{\probdtmc(s,s') \cdot p_{s'}}_{= \, h(s,s')} \quad
\text{for all } s \in S_{>0} \setminus G, 
\end{split}
\end{align}
where $S_{>0}$ and $G \subseteq S$ are as before in Eq.~\eqref{eq:etrencoding}.
Note that the QCQP formulation is exact: its solutions coincide with the solutions to the ETR-formulation.
However, in general, the QCQP is a non-convex optimisation problem --- witnessed by the encoding of the pMC in Fig.~\ref{fig:example-pmc-qcqp} --- and thus not directly amenable to efficient methods (which is a trivial consequence of the ETR-completeness of the problem).

\begin{example}
The ETR constraints for the pMC in Fig.~\ref{fig:example-pmc-qcqp} to reach state $s_3$ with threshold $\leq \lambda$ are:
\[
p_0 \leq \lambda, \quad
\varepsilon \leq x \leq 1{-}\varepsilon, \quad
p_3 = 1, \quad
p_0 \geq x{\cdot}p_1, \quad
p_1 \geq (1{-}x) {\cdot} p_2, \quad
p_2 \geq x {\cdot} p_3.
\]
Quadratic constraints are e.g., $(1{-}x) {\cdot} p_2$ and $x{\cdot}p_1$.
\end{example}


\paragraph{Turning the QCQP into a series of LP problems.}
We present a \emph{sequential-convex-optimisation} (SCP) approach that iteratively solves LPs to find a solution of the QCQP.
This approach is originally presented in~\cite{DBLP:journals/mp/Yuan15,mao2018successive,chen2013optimality} and has been adapted for pMCs in~\cite{DBLP:journals/corr/abs-2107-00108}.  
Eventually, the solution of one of these LPs will provide a solution to the original QCQP. 
Checking whether this is indeed the case can be done by model checking the instantiated pMC. 
Given an assignment $\hat{v}$ to $v$, 
we obtain an affine approximation in the form of a linearisation $\hat{f}$ of the polynomial $f$ around $\hat{v}$:
\begin{equation}
\label{eqn:linearisation}
\hat{f}_{\hat{v}} \ = \ f[\hat{v}] + \nabla f[\hat{v}]^T \cdot (v-\hat{v})
\end{equation}
where $\nabla f[\hat{v}]$ is the gradient of the polynomial $f$ at $\hat{v}$.
Consequently, the LP approximation around $\hat{v}$ of the constraint (\ref{eq:etr-encoding}) then reads:
\[
p_s \ \geq \ \sum_{s' \in S_{>0}} h_{\hat{v}}(s,s')
\quad
\text{for all } s \in S_{>0} \setminus G.
\]
However, the LP from the conjunction of these variables is not necessarily feasible. 
To remedy this, we relax the LP by introducing a penalty variable $k_s$ for each state $s$ that intuitively allows violating the constraints up to some degree. 
By adapting the objective function, we encourage the solver to not violate the constraints, turning (\ref{eq:etr-encoding}) into:
\begin{align}
    \begin{split}
\label{eq:penalties} 
& \text{minimise } p_{\sinit} + \tau {\cdot} \sum_s k_s \text{ s.t.}\quad \\
& k_s + p_s \ \geq \ \sum_{s' \in S_{>0}} h_{\hat{v}}(s,s') \quad
\text{for all } s \in S_{>0} \setminus G. 
\end{split}
\end{align}
The constant $\tau$ is chosen sufficiently large.
To ensure that the linearisation is accurate, we enforce that the solution is sufficiently close to $\hat{v}$ using \emph{trust regions}~\cite{DBLP:journals/mp/Yuan15}.
\begin{example}
After a linearisation around an assignment for the parameter instantiation $\hat{x}$, $\hat{p_0}$ through $\hat{p_2}$, the resulting LP formulation for our example pMC requires to minimise $p_0 + \tau \cdot \sum_{} k_i$ under $k_i \geq 0$, $p_3 = 1$, $p_0 \leq \lambda$, and the linear constraints:
\[
\begin{array}{rcl}
k_0 + p_0 & \ \geq \ & \hat{x}{\cdot}\hat{p_1} + \hat{p_1}{\cdot}(x{-}\hat{x}) + \hat{x}{\cdot}(p_1{-}\hat{p_1}) \\[1ex]
k_1 + p_1 & \geq & p_2 - \hat{x}{\cdot}\hat{p_2} - \hat{p_2}{\cdot}(x{-}\hat{x}) - \hat{x}{\cdot}(p_2{-}\hat{p_2}) \\[1ex]
k_2 + p_2 & \geq & \hat{x}{\cdot}\hat{p_3} + \hat{p_3}{\cdot}(x{-}\hat{x}) + \hat{x}{\cdot}(p_3{-}\hat{p_3}),
\end{array}
\]
and some constraints on the trust region with radius $\delta > 0$, where $\delta' = \delta{+}1$:
\[
\nicefrac{\hat{p_0}}{\delta'} \geq p_0 \geq \hat{p_0}{\cdot}\delta',
\ \
\nicefrac{\hat{p_1}}{\delta'} \geq p_1 \geq \hat{p_1}{\cdot}\delta',
\ \ 
\nicefrac{\hat{p_2}}{\delta'} \geq p_2 \geq \hat{p_2}{\cdot}\delta',
\text{ and }
\nicefrac{\hat{x}}{\delta'} \geq x \geq \hat{x}{\cdot}\delta'.
\]
\end{example}

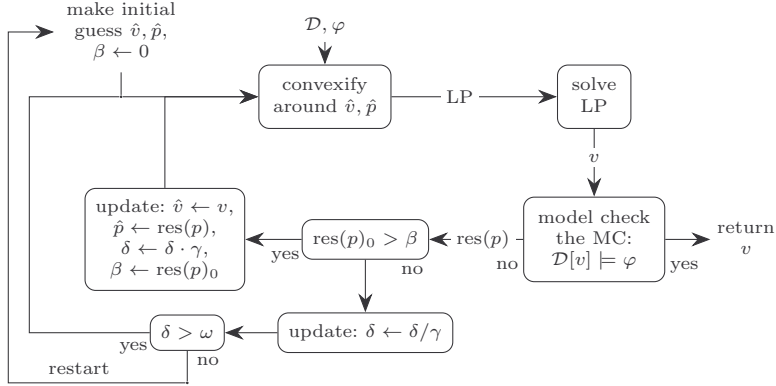
\begin{figure}[t]
\centering
\begin{tikzpicture}[every node/.style={font=\scriptsize},every text node part/.style={align=center},>={Stealth[scale=1.5]}]
\node[draw,rectangle, inner sep=4pt, rounded corners] (conv) {\begin{tabular}{c}convexify\\around $\hat{v}, \hat{p}$\end{tabular}};
\node[circle,left=1.8cm of conv,inner sep=0pt, fill=black] (lconv) {};
\node[above=0.3cm of lconv] (start) {\begin{tabular}{c}make initial \\ guess $\hat{v}, \hat{p}$,\\
	$\beta \leftarrow 0$\end{tabular}};
\node[draw,rectangle, right=2.2cm of conv, inner sep=4pt, rounded corners] (dc) {\begin{tabular}{c}solve \\ LP \end{tabular}};
\node[draw,rectangle, below=0.9cm of dc, inner sep=4pt, rounded corners](check) {\begin{tabular}{c}model check\\ the MC:\\$\pdtmc[v] \models \varphi$\end{tabular}};
\node[draw,rectangle, left=1.25cm of check, inner sep=4pt, rounded corners](checkobj) {$\textrm{res}(p)_0 >\beta$};
\node[draw,rectangle, left= 0.75cm of checkobj,  inner sep=4pt, rounded corners] (update) {update: $\hat{v} \leftarrow v$,\\$\hat{p}\leftarrow \textrm{res}(p),$\\ $\delta\leftarrow\delta\cdot \gamma$,\\$\beta\leftarrow \textrm{res}(p)_0$};
\node[draw,rectangle, below= 0.7cm of checkobj,  inner sep=4pt, rounded corners] (update_rej) {update: $\delta\leftarrow\delta/ \gamma$};
\node[draw,rectangle, left = 0.70cm of update_rej, inner sep=4pt, rounded corners](checktrust) {$\delta >\omega$};
\node[above=0.3cm of conv] (input) {$\pdtmc,\varphi$};
\node[right=0.6cm of check, inner sep=1pt] (solution) {\begin{tabular}{c}return \\$v $\end{tabular}};
\node[circle,below=0.4cm of checktrust,inner sep=0pt, fill=black] (solution_trust-circ) {};
\node[above right=-1.0cm and 0.03cm of update] (res) {};
\draw[->] (check) --  (solution);
\draw[->] (checkobj) --(update_rej);
\draw[->] (check) --  node[pos=0.1,right,below right=-0.02cm and 0.1cm of check] {} (checkobj);
\draw[->] (checkobj) --  node[pos=0.1,right,below right=-0.35cm and -0.4cm of check] {} (update);
\draw[->] (update.north) |- (conv);
\draw[->] (update_rej) -- (checktrust);
\draw [-] (checktrust) -- ++(-2.1,0) |- (lconv);
\draw [-] (checktrust)-- (solution_trust-circ);
\draw[->] (solution_trust-circ) -| node[above,pos=0.3] {restart} ($(start)+(-1.5,0)$) -- (start);
\node[above left=-0.30cm and 0.0cm of checktrust] {};
\node[below right=0.0cm and -0.50cm of checktrust] {};
\draw[->] (lconv) -- (conv);
\draw[-] (start) -- (lconv);
\draw[->] (input) -- (conv);
\draw[->] (conv) -- node[above] {} (dc);
\draw[->] (dc) --  (check);
\node[fill=white,rectangle,inner sep=2pt,below left=-0.77cm and 0.1cm of check,align=center] (j) {$\textrm{res}(p)$};
\node[fill=white,rectangle,below=0.25cm of dc,align=center,inner sep=2pt] (j) {$v$};
\node[fill=white,rectangle,right=0.65cm of conv,align=center,inner sep=2pt] (i) {LP};
\node[fill=none,rectangle,below left=-0.20cm and -0.05cm  of checkobj,align=center,inner sep=2pt] (i3) {yes};
\node[fill=none,rectangle,below right=-0.00cm and -0.45cm of checkobj,align=center,inner sep=2pt] (i4) {no};
\node[fill=none,rectangle,below left=-0.20cm and -0.05cm  of checktrust,align=center,inner sep=2pt] (i3) {yes};
\node[fill=none,rectangle,below right=-0.00cm and -0.45cm of checktrust,align=center,inner sep=2pt] (i4) {no};
\node[fill=none,rectangle,below right=-0.40cm and -0.05cm of check,align=center] (i5) {yes};
\node[fill=none,rectangle,below left=-0.40cm and -0.05cm of check,align=center] (i6) {no};
\end{tikzpicture}
\caption{Feasibility checking using an integrated SCP-PMC approach}
\label{fig:scp-pmc-approach}
\end{figure}

\paragraph{Integration with probabilistic model checking.}
These methods can ensure the correctness of the solution only when the initial point is feasible, which amounts to solving the parameter synthesis problem. Thus, this assumption cannot be made.
Instead, we employ the following scheme. 
A PMC step is integrated into the SCP approach, see details in~\cite{DBLP:journals/corr/abs-2107-00108} to verify whether the instantiated pMC by the solutions for $x$, $v = \textrm{res}(x)$, obtained from the LP improves upon the previous instantiation.
If indeed the computed reachability probability from the initial state (by convention: $\textrm{res}(p)_0$) increases over the result ($\beta$) from the previous iteration, then these values are used in the next iteration.
Otherwise, the radius of the trust region is contracted (by a factor $\gamma < 1$) and the LP is solved again. If the trust region becomes too large ($> \omega$), we may have hit a local optimum and restart.
A schematic overview of the resulting iterative procedure is given in Fig~\ref{fig:scp-pmc-approach}.
Trust region methods converge under regularisation assumptions such as Lipschitz continuity for the gradients of the functions in the objective and constraints.
These conditions are satisfied for our QCQP formulation.

\paragraph{What is different in CCP?}
Above, we outlined the SCP approach from Fig.~\ref{fig:feasibility-spectrum}. The figure also mentions the CCP procedure, which approximates the QCQP using a series of convex programs rather than a series of linear programs. The main trick is to write the quadratic terms as a sum of a convex and a concave part, and then to linearise the concave part. Contrary to the SCP approach, this construction ensures that the constraints all (in some sense) overapproximate the original solution, and thus that (whenever the penalty variables are assigned to zero) a solution to the convexified QCQP is a solution to the original QCQP. 


\paragraph{Experimental evaluation.}
Experiments~\cite{DBLP:journals/corr/abs-2107-00108} on several benchmarks indicate that the SCP-PMC technique can solve feasibility for pMCs with hundreds of thousands of states and tens of thousands of parameters.
Why not use plain SCP?
The integration of PMC techniques yields an improvement of multiple orders of magnitude in run time compared to plain SCP, and guarantees the solution’s correctness. 

\section{Controller Synthesis under Partial Observability}
\label{sec:pomdps}

As indicated in the introduction of this survey, the feasibility problem for parametric MCs --- find a parameter instantiation satisfying reachability objective $\varphi$ --- is equivalent to the finite-memory controller synthesis problem for $\varphi$ in POMDPs.
A POMDP is in fact a finite-state MDP in which to each state a set of observations is associated.
States with the same observations have the same set of enabled actions.

A policy for a POMDP, e.g., to maximise the probability to eventually reach a given state, has to base its decisions on the sequence of observations seen so far. 
This contrasts policies for MDPs that have the sequence of states visited so far at their disposal to make a decision.
Deciding whether there exists an observation-based policy satisfying an infinite-horizon reachability specification $\varphi$ is undecidable~\cite{DBLP:journals/ai/MadaniHC03}. 
Optimal policies need infinite memory.
For computational tractability, policies are therefore often restricted to finite memory and are randomised: an enabled action is taken with some probability.
Let us for simplicity restrict ourselves to memory-less (aka: positional) policies.
These policies have memory size one.

\begin{theorem} \cite{DBLP:conf/uai/Junges0WQWK018}
The decision problem whether there exists a memory-less policy for a POMDP satisfying $\varphi$ is polynomial-time equivalent to the feasibility checking problem for $\varphi$ on a corresponding pMC.
\end{theorem}
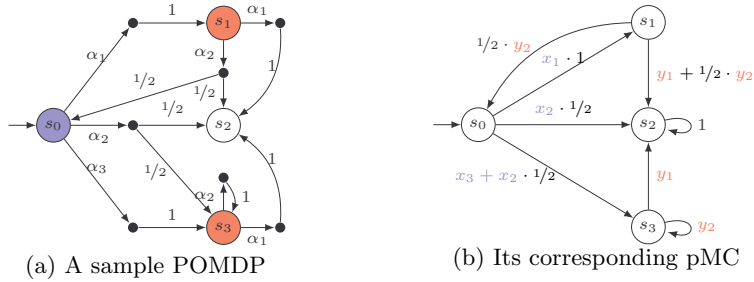
\begin{figure}[tb]
\centering
\begin{subfigure}{0.42\textwidth}
\centering
\scalebox{0.9}{
\begin{tikzpicture}[every node/.style={font=\scriptsize},st/.style={circle, inner sep=2pt, draw}]
\node[st, fill=blue!40, initial, initial text=] (s0) {$s_0$};
\node[st, right=2cm of s0, fill=white!40] (s2)  {$s_2$};
\node[st, above=of s2, fill=red!60] (s1)  {$s_1$};
\node[st, below=of s2, fill=red!60] (s3)  {$s_3$};

\node[circle, inner sep=1.5pt, fill=black, left=1cm of s1] (a1) {};
\node[circle, inner sep=1.5pt, fill=black, left=1cm of s2] (a2) {};
\node[circle, inner sep=1.5pt, fill=black, left=1cm of s3] (a3) {};
\node[circle, inner sep=1.5pt, fill=black, right=0.5cm of s1] (a4) {};
\node[circle, inner sep=1.5pt, fill=black, below=0.4cm of s1] (a5) {};
\node[circle, inner sep=1.5pt, fill=black, right=0.5cm of s3] (a6) {};
\node[circle, inner sep=1.5pt, fill=black, above=0.4cm of s3] (a7) {};

\draw[->] (s0) -- node[above] {$\act_1$} (a1);
\draw[->] (s0) -- node[below] {$\act_2$} (a2);
\draw[->] (s0) -- node[above] {$\act_3$} (a3);
\draw[->] (s1) -- node[above] {$\act_1$} (a4);
\draw[->] (s1) -- node[left] {$\act_2$} (a5);
\draw[->] (s3) -- node[below] {$\act_1$} (a6);
\draw[->] (s3) -- node[left] {$\act_2$} (a7);

\draw[->] (a1) -- node[above] {$1$} (s1);
\draw[->] (a2) -- node[above] {$\nicefrac{1}{2}$} (s2);
\draw[->] (a2) -- node[left] {$\nicefrac{1}{2}$} (s3);
\draw[->] (a3) -- node[above] {$1$} (s3);

\draw[->] (a4) edge[bend left] node[above] {$1$} (s2);

\draw[->] (a5) -- node[above] {$\nicefrac{1}{2}$} (s0);
\draw[->] (a5) -- node[left] {$\nicefrac{1}{2}$} (s2);

\draw[->] (a6) edge[bend right] node[above] {$1$} (s2);

\draw[->] (a7) edge[bend left] node[right] {$1$} (s3);
\end{tikzpicture}
}
\vspace*{-0.2cm}
\subcaption{A sample POMDP}
\label{fig:POMDPtoAnyPMC:POMDP}
\end{subfigure}  
\hspace{2em}
\begin{subfigure}{0.42\textwidth}
\centering
\scalebox{0.9}{
\begin{tikzpicture}[every node/.style={font=\scriptsize},st/.style={circle, inner sep=2pt, draw}]
\node[st,initial,initial text=] (s0) {$s_0$};
\node[st, right=2cm of s0] (s2)  {$s_2$};
\node[st, above=of s2] (s1)  {$s_1$};
\node[st, below=of s2] (s3)  {$s_3$};

\draw[->] (s0) -- node[above] {${\color{blue!40}x_1} \cdot 1$} (s1);
\draw[->] (s0) -- node[above] {${\color{blue!40}x_2} \cdot \nicefrac{1}{2}$} (s2);
\draw[->] (s0) -- node[left, align=center] {${\color{blue!40}x_3 + x_2} \cdot \nicefrac{1}{2}  $} (s3);

\draw[->] (s3) edge[loop right] node[right] {$\color{red!60}y_2$} (s3);
\draw[->] (s3) edge node[right] {$\color{red!60}y_1$} (s2);
\draw[->] (s1) edge[bend right] node[left] {$\nicefrac{1}{2}\cdot {\color{red!60}y_2}$\ \ } (s0);
\draw[->] (s1) edge node[right] {$ {\color{red!60}y_1} + \nicefrac{1}{2}\cdot {\color{red!60}y_2}$} (s2);
\draw[->] (s2) edge[loop right] node[right] {$1$} (s2);
\end{tikzpicture}
}
\vspace*{-0.2cm}
\subcaption{Its corresponding pMC}
\label{fig:POMDPtoAnyPMC:PMC}
\end{subfigure} 
\vspace*{-0.3cm}
\caption{The connection between POMDPs and pMCs}
\end{figure}
Together with Theorem~\ref{th:etr-complete}, this yields that finding a policy for a POMDP satisfying $\varphi$ is ETR-complete.
This improves the known complexity results for the POMDP controller synthesis decision problem: being NP-hard, SQRT-SUM-hard and being in PSPACE~\cite{DBLP:journals/toct/VlassisLB12}.
How to obtain a corresponding pMC?
One can obtain a pMC for a POMDP under an arbitrary randomised memory-less policy in a simple way.
The topology of the pMC and POMDP are identical.
The parametric transition probabilities in the pMC are obtained from the (unknown) probabilities to select actions in the POMDP by an arbitrary randomised policy. 
The following example illustrates this for a memory-less policy.
\begin{example}
Fig.~\ref{fig:POMDPtoAnyPMC:POMDP} depicts a sample POMDP where colors indicate the observations at a state, and actions are indicated by $\alpha_i$.
Now consider the following memory-less randomised policy.
In any red colored state, it takes action $\act_1$ with probability $y_1$, and action $\act_2$ with probability $y_2$. 
At any blue colored state, action $\act_i$ is taken with probability $x_i$. 
This perspective directly yields the pMC of Fig.~\ref{fig:POMDPtoAnyPMC:PMC}.
\end{example}

\noindent
The connection and construction for memory-less policies can be extended to finite-memory controllers.
We employ \textit{randomised finite-state controllers (FSCs)}~\cite{poupart2004bounded}, that are essentially extensions of Moore machines.
A POMDP satisfies $\varphi$ under some $k$-FSC (an FSC with $k$ memory states) if and only if a POMDP that is obtained by a $k$-\emph{unfolding} of the original POMDP satisfies $\varphi$ under a memory-less policy.
This unfolding can be viewed as the synchronous product construction of the POMDP and the $k$-FSC.
The induced pMC of the unfolded POMDP has a state space that is linear in $k$ and its number of parameters is quadratic in $k$ and linear in the total number of observations as well as the maximal number of actions over all observations.

\begin{figure}[h]
\centering
\includegraphics[scale=0.3]{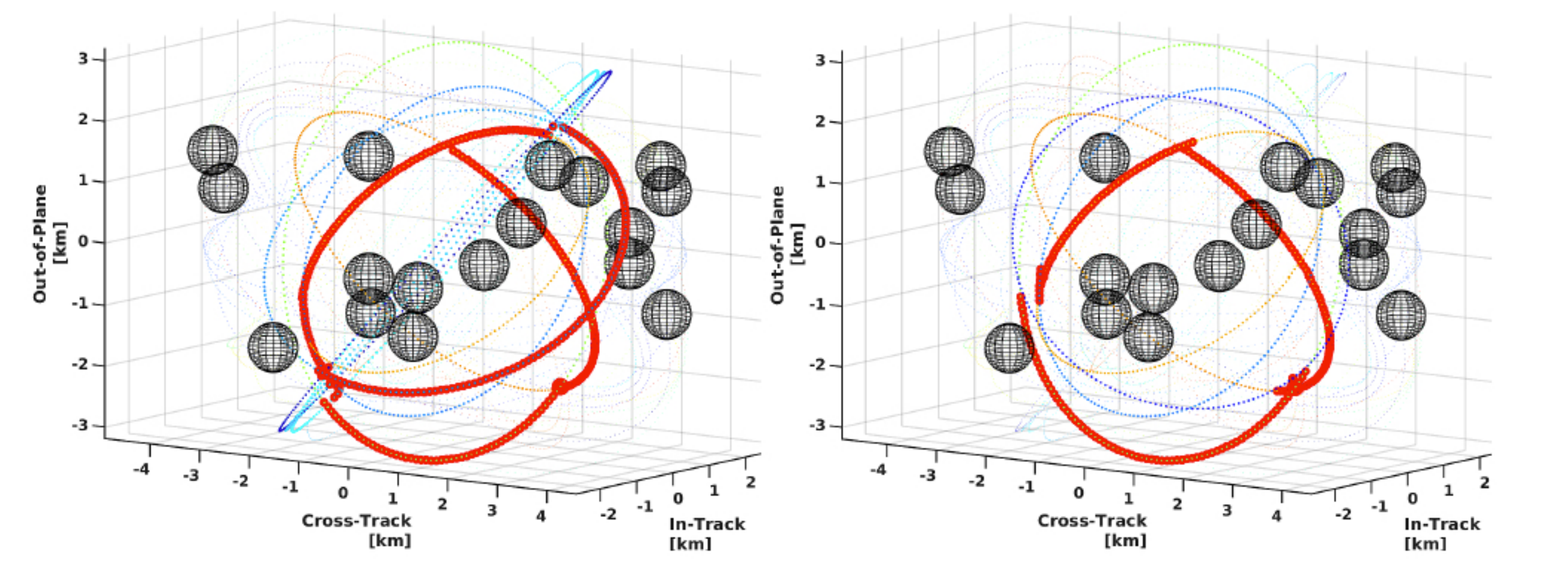}
\vspace*{-0.3cm}
\caption{Satellite trajectories obtained by pMC feasibility checking}
\label{fig:satellite-scp}
\end{figure}

\paragraph{A satellite collision avoidance example.}
We consider a swarm of satellites following circular orbits with a radius of 7728 km.
A satellite can follow its current trajectory with no fuel usage.  
A trajectory is discretised in numerous (7200) time steps.
Based on regularly obtained observations, at every observation the satellite can either stay in its current trajectory or can switch to a close ($\leq$ 250 km apart) trajectory.
In total 36 trajectories are possible.
The objective is to compute a switching policy that ensures the satellite to avoid collisions with other objects with high (0.995) probability.
The POMDP for this scenario ranges from 6K states and a few hundred observations for $k{=}0$ (memory-less) to 288K states and thousands of observations for $k{=}5$.
The feasibility of the resulting pMC is analysed using the SCP-PMC approach from Section~\ref{sec:feasibility}.
Figure~\ref{fig:satellite-scp} indicates the obtained policies for a memory-less policy (left) and a policy with memory size five (right) obtained in about 20 seconds and five minutes of computation time respectively. 
The initial satellite position is indicated by a fat red circle.
The black spheres represent the obstacles to be avoided.
The average length of the trajectory with the finite-memory policy is about half the length for the for the memory-less policy. 

\section{Epilogue}

Finally, we present some interesting research directions on parameter synthesis along with some first results in these directions. 

\subsubsection*{Monotonicity checking.} 
It is often the case that perturbing parameter values has a monotonic influence on the induced reachability probability. 
Consider, e.g., a network protocol where increasing the channel quality typically decreases the protocol's packet drop rate. 
Increasing parameter values that encode the reliability of hardware components will typically increase the system's reliability. 
The decision problem whether the reachability probability function is monotonic in a given parameter is however ETR-hard~\cite{DBLP:conf/atva/SpelJK19}. 
Nevertheless, checking sufficient properties on the topology of the pMC yields effective methods to establish monotonicity~\cite{DBLP:conf/atva/SpelJK19}. 
Whenever monotonicity (in some parameter) is established, region verification and feasibility checking become significantly cheaper. 
In general, however, the overhead of first establishing monotonicity does not always pay off. 

Recently, \cite{DBLP:conf/tacas/SpelJK21} intertwined the region verification method outlined in Section~\ref{sec:region} with establishing monotonicity (within that region). 
The main benefit is that when the over-approximation of the region verification is too coarse, the obtained bounds from the verification help to establish monotonicity whereas on the other hand monotonicity results simplify the refinement loop in region verification. 
This yields an effective method to solve the \emph{optimal}\footnote{In fact, $\varepsilon$-optimal feasibility, i.e., finding a parameter instantiation that achieves the maximal or minimal reachability probability up to a given accuracy $\varepsilon > 0$.} feasibility problem.

When assuming monotonicity, hill climbing (gradient descent, GD) methods for feasibility checking perform especially well: the optimal value will be in a corner of the parameter space and the GD methods will accelerate sampling towards that corner. 
We furthermore observe that the performance here neither requires establishing monotonicity a priori, nor does it require monotonicity in all parameters. 
The key step to enable GD methods is to efficiently compute the gradient. 
The recent paper~\cite{Speletal2021} shows that this is possible and additionally empirically compares various GD methods in the context of feasibility checking. 


\subsubsection*{Transfer to other models.} 
Various approaches in this survey carry over to parametric MDPs\footnote{Solution functions for pMDPs are challenging, but can be thought of as the maximum over the solution functions for the induced pMCs.}.
This applies e.g., to complexity results for feasibility, parameter lifting, and mathematical optimisation for feasibility.
Various works consider even richer models. 
In particular, the methods described here can be extended towards parametric probabilistic timed automata~\cite{DBLP:conf/qest/HartmannsKKS21} and to controller synthesis for uncertain POMDPs, see below. 
Similarly, there exist various approaches for parametric continuous-time MCs, see, e.g.,~\cite{DBLP:conf/cmsb/CeskaDKP14,DBLP:journals/jss/CalinescuCGKP18,DBLP:journals/pe/GoubermanST19} and parameter synthesis has been applied to stochastic population models~\cite{DBLP:conf/cmsb/HajnalNPS19} and to accelerate solving hierarchical MDPs~\cite{DBLP:journals/corr/abs-2206-02653}.

Beyond Markov models, probabilistic graphical models in general and Bayesian networks in particular are widespread to describe complex conditional probability distributions.
Recent work~\cite{DBLP:journals/corr/abs-2105-14371,salmanikatoen2022} shows that ideas and methods for parameter synthesis in Markov chains as described in this survey significantly improve upon existing methods for parametric Bayesian networks~\cite{DBLP:journals/jair/ChanD02}. 
Vice versa, some inference techniques do yield interesting alternatives for the analysis of (finite-horizon properties in) pMCs~\cite{DBLP:conf/cav/HoltzenJVMSB20}. 

Synthesis techniques explained in this survey are also used for stochastic hybrid systems.
The approach advocated in~\cite{DBLP:conf/formats/PeruffoA21} takes a parametric stochastic differential equation (SDE) and a specification $\varphi$, computes a parametric formal abstraction, yielding a (parametric) abstract model, and synthesises
parameters on the abstract model in order to satisfy $\varphi$ for the SDE. 
The abstraction can be iteratively refined to increase its precision. 

Parameter synthesis has recently also been considered for the richer type of probabilistic hyper-properties~\cite{DBLP:conf/lpar/AbrahamBBD20}, properties that are relevant in e.g., security.

\subsubsection*{Topology synthesis.} 
The methods presented here are tailored towards continuous parameter values with a continuous solution function (which means that most methods assume graph-preserving regions or acyclic pMCs). 
A related setting studies discrete parameters and varying topologies. 
Such settings naturally occur when synthesising implementations of probabilistic programs and controllers in probabilistic settings. 
In a nutshell, methods rely either on abstraction-refinement~\cite{DBLP:conf/tacas/CeskaJJK19}, similar to Section~\ref{sec:region}, or on inductive synthesis~\cite{DBLP:conf/fm/CeskaHJK19} inspired by ideas such as programming-by-example and program synthesis.
These two methods can be efficiently combined~\cite{DBLP:conf/tacas/Andriushchenko021}. 
Despite a slightly different focus, the integration of the discrete and continuous setting seems a natural next step. 

\subsubsection*{Variations.} 
Beyond the problem statements discussed in this paper, there are various related problem statements.

\paragraph{Robust policies.}
For parametric MDPs, one can either quantify:
\begin{itemize}
    \item (\emph{robust policies}) first over the policies and then over the parameters, or
    \item (\emph{robust parameters}) first over the parameters and then over the policies.
\end{itemize}
The generalisation of the approaches in this paper consider the latter. 
While some ideas carry over, there are some differences. 
Furthermore, finding memory-less robust policies is theoretically harder than finding robust parameters and, in general, optimal robust policies require memory~\cite{DBLP:journals/jcss/JungesK0W21,DBLP:conf/qest/ArmingBCKS18}. 
We remark that the notion of robustness itself can also be considered on pMCs, using, e.g., perturbation analysis~\cite{DBLP:conf/concur/ChenFRS14}.

\paragraph{Interval and other uncertainty models.}
Many models assume a more local notion of uncertainty in transition probabilities. 
Typically, the verification problem for such models aims at robustness against all probability distributions within the uncertainty sets; this may also mean robust policies as discussed above.
The transitions of \textit{uncertain MCs or MDPs} are equipped with, e.g., \emph{probability intervals}, \emph{likelihood functions}, or form even more general uncertainty sets~\cite{DBLP:journals/tcs/BartDFLMT18,seshia_et_al_cav_13,DBLP:journals/ai/GivanLD00,DBLP:conf/cdc/WolffTM12,nilim2005robust,DBLP:conf/isipta/KrakTB19}. 
For a detailed handling of different types of uncertainty sets, we refer to~\cite{wiesemann2013robust}.
Extensions to uncertain POMDPs also exist~\cite{DBLP:conf/ijcai/Suilen0CT20,DBLP:conf/aaai/Cubuktepe0JMST21}.

\paragraph{Distributions over parameters.}
Rather than assuming the best or worst case for parameter values, one can equip parameters with a parameter distribution~\cite{DBLP:conf/qest/ArmingBCKS18}. 
For pMCs, this assumption then enables sampling-based approaches that yield probably-approximately-correctness (PAC) style guarantees~\cite{DBLP:journals/corr/abs-2112-13020}.
Finally, one may integrate Bayesian updates on parameter distributions with the analysis of pMCs~\cite{DBLP:conf/qest/PolgreenWHA17}. For parametric continous-time MCs, both scenario-based approaches~\cite{DBLP:journals/corr/abs-2205-08300} and a variety of approaches around Gaussian processes exist~\cite{DBLP:journals/iandc/BortolussiMS16,DBLP:conf/tacas/BortolussiS18}.

\paragraph{Model repair and optimisation.}
Parameter synthesis plays an important role in model repair~\cite{DBLP:conf/tacas/BartocciGKRS11,DBLP:conf/nfm/PathakAJTK15}: how can the transition probabilities in a given parameter-less MC that refutes a given specification $\varphi$ be changed such that an MC is obtained satisfying $\varphi$?
Model repair reduces to a feasibility checking problem.
To deal with large state spaces, model repair has been combined with abstraction~\cite{DBLP:journals/iandc/Chatzieleftheriou18}.
Optimal feasibility has been applied to find the optimal bias of coin flips used in self-stabilisation algorithms for distributed systems~\cite{DBLP:journals/dc/VolkBKA22}.

\subsection*{Acknowledgements}
We thank all our co-authors in the work(s) surveyed in this paper: Erika \'Abrah\'am, Christel~Baier, Bernd~Becker, Harold~Bruintjes,  Florian~Corzilius, Murat~Cubuktepe, Christian~Hensel, Lisa~Hutschenreiter, Joachim~Klein, Guillermo~P\'erez, Tim~Quatmann, Ufuk~Topcu, Matthias~Volk, Ralf~Wimmer, Tobias~Winkler and Leonore~Winterer. 
We thank Bahare Salmani, Jip Spel, and Tobias Winkler as well as the reviewers of this volume for their thoughtful feedback on a draft version of this survey.

\bibliographystyle{splncs04}
\bibliography{main}

\end{document}